\documentclass[11pt]{article}

\usepackage[final]{acl}

\usepackage{times}
\usepackage{latexsym}

\usepackage[T1]{fontenc}

\usepackage[utf8]{inputenc}

\usepackage{microtype}

\usepackage{inconsolata}

\usepackage{graphicx}

\usepackage{booktabs} 
\usepackage{hyperref}

\usepackage{amsmath}
\usepackage{amssymb}
\usepackage{mathtools}
\usepackage{amsthm}
\usepackage[capitalize,noabbrev]{cleveref}

\theoremstyle{plain}

\theoremstyle{definition}

\theoremstyle{remark}

\usepackage[textsize=tiny]{todonotes}

\usepackage[utf8]{inputenc} 
\usepackage[T1]{fontenc}    
\usepackage{url}            
\usepackage{amsfonts}       
\usepackage{nicefrac}       
\usepackage{natbib}
\usepackage{float}
\usepackage{xurl} 
\usepackage{threeparttable}
\usepackage{multirow}
\usepackage{fontawesome}
\usepackage{makecell}
\usepackage{tabularx}
\newcolumntype{Y}{>{\centering\arraybackslash}X}
\usepackage{enumitem}
\usepackage{mdframed}
\usepackage{listings}
\usepackage{pifont}
\usepackage{seqsplit}
\usepackage{subcaption}
\usepackage{xcolor}
\usepackage{calligra}
\newcommand{\flame}{\includegraphics[height=1em]{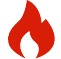}}
\newcommand{\snowflake}{\includegraphics[height=1em]{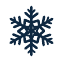}}

\definecolor{neworange}{RGB}{246, 198, 172}
\definecolor{newgreen}{RGB}{72, 212, 90}
\usepackage[most]{tcolorbox}
\newtcolorbox[auto counter, number within=section]{mybox}[2][]{%
  breakable,
  boxrule=0.4pt,
  colback=gray!10!white,
  colframe=gray!60!black,
  fonttitle=\small\bfseries,
  title=Box \thetcbcounter: #2, 
  #1
}
\newcommand\blfootnote[1]{%
  \begingroup
  \renewcommand\thefootnote{}\footnote{#1}%
  \addtocounter{footnote}{-1}%
  \endgroup
}
\title{STELLA: A Multimodal LLM for Protein Functional Annotation via Unified Sequence-Structure Encoding}

\author{
  \textbf{Hongwang Xiao}\textsuperscript{1,2,$\mathsection$,*},
  \textbf{Wenjun Lin}\textsuperscript{1,*},
  \textbf{Xi Chen}\textsuperscript{1},
  \textbf{Hui Wang}\textsuperscript{1},
  \textbf{Kai Chen}\textsuperscript{1}, \\
  \textbf{Jiashan Li}\textsuperscript{1,3},
  \textbf{Yuancheng Sun}\textsuperscript{1,4,5},
  \textbf{Sicheng Dai}\textsuperscript{1,4,5},
  \textbf{Boya Wu}\textsuperscript{1},
  \textbf{Qiwei Ye}\textsuperscript{1,$\dagger$} \\[0.5ex]
  \textsuperscript{1}Beijing Academy of Artificial Intelligence \\
  \textsuperscript{2}State Key Laboratory of Multimedia Information Processing, Peking University \\
  \textsuperscript{3}Renmin University of China \quad
  \textsuperscript{4}University of Chinese Academy of Sciences \\
  \textsuperscript{5}Institute of Automation, Chinese Academy of Sciences \\[0.3ex]
}

\begin{document}
\maketitle
\blfootnote{\hspace*{-1.8em}$^{\mathsection}$Project Lead \quad $^{*}$Equal Contribution \\ 
$^{\dagger}$Corresponding Author: \texttt{qwye@baai.ac.cn} \\
\texttt{\{hwxiao, wjlin, chenxi, wanghui, bywu\}@baai.ac.cn} \\
\texttt{chenkai.cn@hotmail.com,} \texttt{lijiashan@ruc.edu.cn} \\ 
\texttt{\{sunyuancheng2021, daisicheng2023\}@ia.ac.cn}}

\begin{abstract}
Understanding the intricate interplay among sequence, structure, and function remains a fundamental challenge in proteomics. The sequence-structure-function paradigm posits that biological roles are governed by the tertiary geometric conformations encoded within primary sequences; consequently, integrating these multi-modal descriptors is imperative for accurate functional annotation. While protein language models (pLMs) have achieved significant progress via representation learning on massive sequence data, they often lack the capacity to incorporate high-resolution structural information and the rich textual context that characterizes protein roles. In this work, we present STELLA, a multimodal LLM that synergistically aligns bimodal (sequence-structure) representations with the textual modality to advance protein functional annotation. By leveraging ESM3 for unified bimodal encoding and Llama-3.1-8B-Instruct for natural language modeling, STELLA achieves state-of-the-art performance in two critical tasks: Functional Description Prediction and Enzyme-catalyzed Reaction Prediction. This study demonstrates that multimodal LLMs represent a paradigm shift beyond pure pLMs, offering a new frontier for protein biology and biomedical discovery. The codes can be accessed via \url{https://github.com/ocx-lab/STELLA}.
\end{abstract}

\section{\textbf{Introduction}}\label{intro}
Protein biology centers on the intricate interplay among three fundamental modalities: sequence, structure, and function. The central tenet of structural biology—that sequences dictate structures, which in turn govern functions—underscores the deterministic relationship between a protein’s primary sequence and its biological function. Specifically, the tertiary topology of a protein defines its interaction landscape with ligands, substrates, or inhibitors, thereby mediating essential activities such as enzymatic catalysis and molecular recognition. Deciphering these functional mechanisms is paramount for elucidating disease pathology—where protein dysfunctions are frequently the primary drivers—and for accelerating drug discovery, metabolic engineering, and the design of novel biocatalysts for industrial biotechnology.

Recent decades have witnessed an explosion of structural data, characterized by the growth of experimentally solved structures in the RCSB Protein Data Bank (PDB)\footnote{\url{https://www.rcsb.org/}}~\citep{10.1093/nar/28.1.235} and the vast repository of high-confidence predictions in the AlphaFold Database (AFDB)\footnote{\url{https://alphafold.ebi.ac.uk/}}~\citep{10.1093/nar/gkab1061}. While advancements in structural proteomics (e.g., AlphaFold 2~\citep{Jumper2021HighlyAP}, AlphaFold3~\citep{abramson2024accurate}, Chai-1~\citep{chai2024chai}, Boltz-1~\citep{wohlwend2025boltz} and OpenComplex2~\citep{opencomplex2025towards} have reached unprecedented levels of accuracy, our understanding of protein functions has not kept pace with this structural revolution. A profound functional annotation gap remains, as the biological roles of the majority of sequenced and folded proteins remain elusive. Consequently, despite the maturation of structure prediction, the challenge has shifted toward harnessing these structural insights to facilitate functional elucidation—a task that requires moving beyond static geometry to understand complex biological processes, subcellular dynamics, and context-dependent activities.

To bridge this gap, protein language models (pLMs) have been developed to learn joint sequence-structure representations~\citep{su2023saprot, 10.1093/bib/bbaf120}. However, while these models excel at capturing biophysical attributes, they often struggle to integrate the textual modality—the descriptive knowledge essential for defining biological function. Emerging multimodal large language models (MLLMs), such as Prot2Text~\citep{abdine2023prot2text}, ProteinGPT~\citep{xiao2024proteingpt}, ProtChatGPT~\citep{wang2024protchatgpt}, ProteinChat~\citep{Huo2024.08.19.608729}, have attempted to integrate protein data with natural language. Nevertheless, these frameworks typically rely on disparate pre-trained encoders for sequence and structure, necessitating complex fusion layers that increase computational overhead and complicate gradient-based optimization. This architectural fragmentation motivates our investigation into ESM3~\citep{hayes2024simulating}, a frontier pLM, as a unified protein encoder. By embedding sequence and structure into a cohesive latent space, ESM3 offers a streamlined yet potent foundation for multimodal integration within an LLM.

In this work, we present STELLA, a multimodal LLM designed to synergize sequence-structure with natural language (function). STELLA integrates the esm3\_sm\_open\_v1 (1.4B) encoder with the Llama-3.1-8B-Instruct model~\citep{dubey2024llama}, establishing a new paradigm that leverages the unified encoding capacity of pLMs and the superior generation power of generative LLMs. We demonstrate STELLA's efficacy on two critical tasks: \textbf{Functional Description Prediction (\texttt{FP})} and \textbf{Enzyme-catalyzed Reaction Prediction (\texttt{EP})}, representing the \textbf{global and biochemical dimensions of protein functionality}, respectively. While \texttt{FP} requires the generation of comprehensive narratives describing biological roles (e.g., signal transduction or DNA repair), \texttt{EP} demands the precise identification of catalytic specificity. STELLA achieves state-of-the-art performance across both tasks, underscoring the potential of multimodal LLMs to serve as a transformative tool for protein biology and biomedical discovery, transcending the limitations of traditional pLMs. Contributions include:




\textbf{1}. We develop STELLA, a multimodal LLM that simplifies protein-to-text integration via a unified bimodal encoder, setting a new state-of-the-art benchmark for \texttt{FP} and \texttt{EP}.

\textbf{2}. We release OPI-Struc, a large-scale multimodal instruction-tuning dataset that bridges the gap between protein structures and natural language annotations, providing a foundational resource for the protein-LLM community.

\textbf{3}. We establish an innovative paradigm in computational protein science, demonstrating that multimodal LLMs can synergize with pLM-based representations to achieve high-fidelity, context-aware functional characterization.
\section{\textbf{Related Work}}\label{sec:related_work}

\subsection{\textbf{Protein Representation Learning}}
Protein representation learning seeks to develop models capable of extracting biologically meaningful features from diverse modalities, including sequences, structures, and functions. Early foundations were laid by unimodal sequence models such as ProtBERT~\citep{9477085}, ESM-2~\citep{lin2023evolutionary}, and ProtGPT2~\citep{Ferruz2022ProtGPT2IA}, which capture the "grammar" of amino acids through large-scale pre-training. Expanding beyond sequences, research has branched into two primary trajectories. One trajectory focuses on cross-modal alignment between sequences and natural language; for instance, ProtST~\citep{xu2023protst} employs contrastive learning to align protein representations with textual descriptors, while ProteinDT~\citep{liu2023text} leverages text-conditioned diffusion for protein design. Another one emphasizes structural integration, exemplified by SaProt~\citep{su2023saprot}, which pioneered a structure-aware vocabulary by encoding residues and Foldseek-based tokens into a unified ESM architecture. Despite these multi-modal efforts, these models are typically optimized for latent space alignment or specific design tasks, often lacking the generative reasoning capacity required for nuanced functional characterization.

\subsection{\textbf{LLMs for Protein Biology}}
The emergence of Large Language Models (LLMs) has provided a powerful framework for integrating protein-specific data with the extensive "world knowledge" embedded in natural language models. Recent studies have signaled the transformative potential of LLMs in proteomics. Prot2Text first proposed an encoder-decoder architecture to align protein structures with functional narratives by combining ESM-2 and GPT-2. To facilitate biological reasoning, BioMedGPT~\citep{luo2023biomedgpt} and InstructProtein~\citep{wang2023instructprotein} connected sequence encoders with Llama-based models for protein-text generation and question-answering. To incorporate structural insights, ProteinGPT~\citep{xiao2024proteingpt} and ProtChatGPT~\citep{wang2024protchatgpt} utilized specialized encoders such as GVP-GNN and ESM-IF1 to capture tertiary topologies. More recently, ProteinChat~\citep{Huo2024.08.19.608729} integrated the xTrimoPGLM~\citep{chen2024xtrimopglm}  encoder with Vicuna-13B~\citep{zheng2023judging} to enable dialogue-based functional prediction, though it relies solely on sequence inputs. Prot2Chat~\citep{Wang2025Prot2ChatPL} further extended this by incorporating both sequence and structure through LoRA-based fine-tuning. A common characteristic of these contemporary approaches is the reliance on disparate pre-trained encoders for different modalities, which often introduces architectural redundancy and complicates cross-modal optimization. This sets the stage for our investigation into a more streamlined, unified encoding paradigm.

\section{\textbf{Methodology of STELLA}}
\subsection{\textbf{Model Architecture}}
\begin{figure}[!ht]
    \centering
    \includegraphics[width=0.5\textwidth]{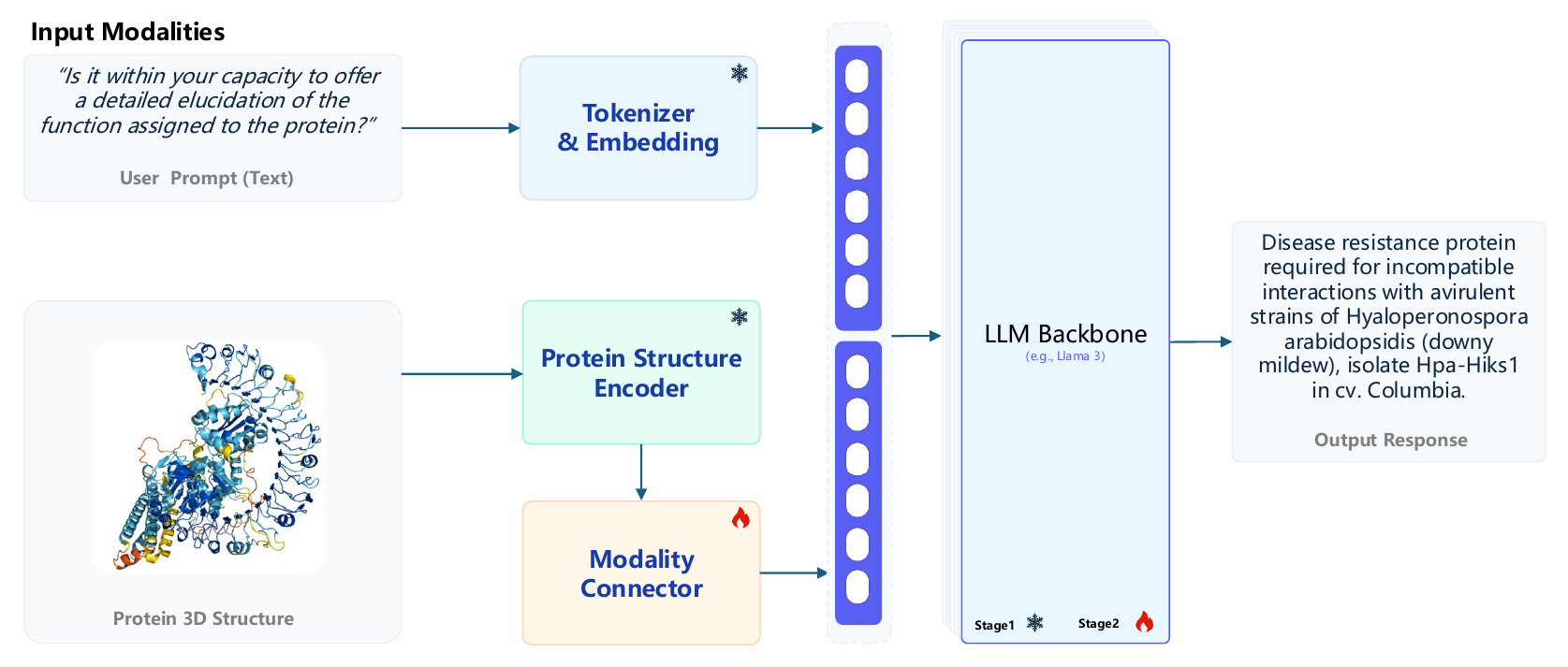}
    \vspace{-0.2in}
    \caption{\textbf{The STELLA architecture and MMIT workflow.} (a) \textbf{Stage 1:} Fine-tuning the modality connector on OPI-Struc (protein encoder and LLM frozen). (b) \textbf{Stage 2:} Joint fine-tuning of the connector and LLM with distinct learning rates, while maintaining a frozen protein encoder. Trainable (\flame) and frozen (\snowflake) modules are indicated. Protein structure credits: AFDB.}
    \label{fig:STELLA_arch}
    \vspace{-0.2in}
\end{figure}

\textbf{Overview.} The architectural design of STELLA is inspired by the LLaVA framework\citep{liu2023visual}, adapting the established vision-language paradigm to the domain of protein biology. As illustrated in Figure~\ref{fig:STELLA_arch}, STELLA comprises three core modules: a \textbf{protein structure encoder}, a \textbf{modality connector}, and a \textbf{LLM}. Following the prevailing training methodology in MLLMs~\citep{he2024efficientmultimodallearningdatacentric}, STELLA undergoes a two-stage Multimodal Instruction Tuning (MMIT) process. Notably, diverging from the standard LLaVA approach—which typically uses distinct datasets for alignment and tuning—STELLA utilizes the same curated dataset for both stages. This strategy is necessitated by the acute scarcity of high-quality protein-text instruction pairs, ensuring maximal data efficiency for both modality alignment and functional reasoning. Detailed prompt templates and hyperparameter configurations are documented in Appendix~\ref{apx:prompt_template_training} and Table~\ref{tab:hyperpara_train_test} of Appendix~\ref{apx:hyperpara_train_test}, respectively.


\textbf{Protein structure encoder.} The protein structure encoder is tasked with projecting tertiary topologies into high-dimensional latent representations. In this study, we employ ESM3, a frontier multimodal protein language model pretrained on a vast repertoire of sequence, structure, and functional descriptors. ESM3 conceptualizes these disparate modalities as discrete token tracks, which are integrated into a unified embedding space via a transformer-based architecture. A pivotal feature of ESM3 is the incorporation of geometric attention within its initial transformer blocks, which provides a strong inductive bias for capturing fine-grained atomic-level spatial details—an essential requirement for accurate functional inference.

\textbf{Modality connector.} The modality connector serves as the neural bridge that aligns the protein-centric representations with the textual latent space of the LLM. In our implementation, we utilize a single linear projection layer as the adapter. Despite its structural simplicity, this linear bottleneck has demonstrated remarkable efficacy in established multimodal frameworks~\citep{liu2023visual, he2024efficientmultimodallearningdatacentric}, offering a computationally efficient yet robust mechanism for cross-modal feature mapping without compromising representational fidelity.

\textbf{LLM.} For the generative and reasoning backbone of STELLA, we employ Llama-3.1-8B-Instruct which represents the state-of-the-art in open-source LLMs, exhibiting superior performance across a diverse spectrum of benchmarks, including general knowledge~\citep{hendrycks2021measuringmassivemultitasklanguage}, mathematical reasoning~\citep{cobbe2021trainingverifierssolvemath}, code generation~\citep{chen2021evaluatinglargelanguagemodels}, tool-use~\citep{berkeley-function-calling-leaderboard,srinivasan2023nexusraven}, long context tasks~\citep{zhang2024inftybenchextendinglongcontext} and multilingual ability~\citep{shi2022languagemodelsmultilingualchainofthought}. Its sophisticated instruction-following capability ensures that the model can interpret nuanced biological prompts, facilitating high-fidelity functional annotation. Furthermore, the integration of Llama Guard 3 ensures the model maintains rigorous safety and reliability standards during text generation.


\subsection{\textbf{Multimodal Instruction Dataset}}
\begin{figure}[!htbp]
    \vspace{-0.1in}
    \centering
    \includegraphics[width=0.9\linewidth]{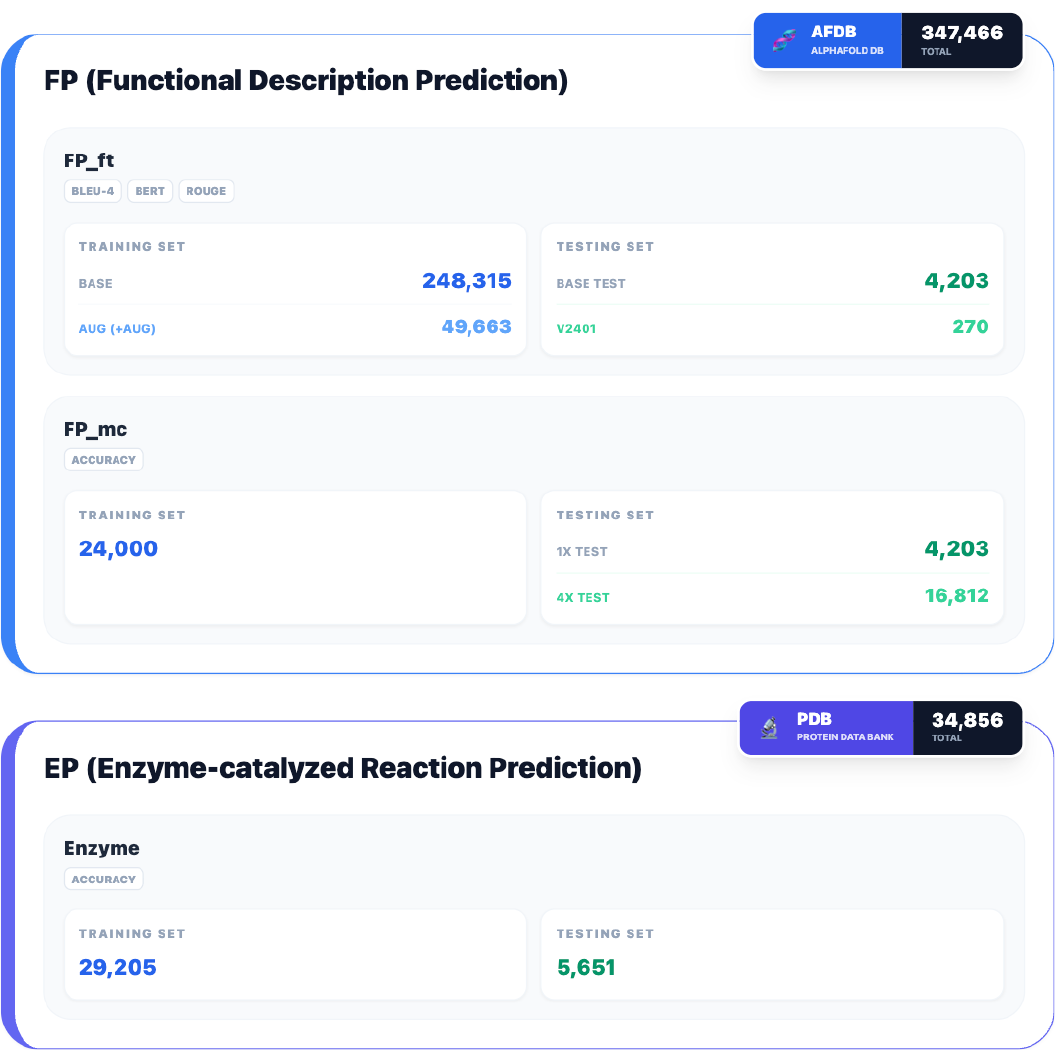}
    \caption{\textbf{OPI-Struc dataset statistics.} The \texttt{FP} task includes \texttt{FP\_ft} (evaluated on 2022\_04 hold-out and 2024\_01 OOD sets) and \texttt{FP\_mc} (with/without option permutation). Example instructions are provided in Appendix Box~\ref{box:ex_func_mc_train}--\ref{box:ex_func_ft_train}.}
    \label{fig:dataset_stats}
    \vspace{-0.1in}
\end{figure}

\textbf{Data overview.} To facilitate MMIT, we curated the Open Protein Instructions for Structures (OPI-Struc) dataset, which synergistically integrates protein structural modalities with descriptive natural language. It differs with prior sequence-only (Mol-Instructions \citep{fangmol}) or property-prediction (PEER \citep{xu2022peer}) datasets. Aligned with the \texttt{FP} and \texttt{EP} tasks, OPI-Struc is stratified into two primary domains: \texttt{\textbf{Function}} and \texttt{\textbf{Enzyme}} (representative examples are provided in Appendix~\ref{apx:examples_instruction_data}). Within the \texttt{Function} domain, samples are further categorized based on their linguistic format: \texttt{Func\_ft}, comprising free-text question-answer pairs for generative evaluation, and \texttt{Func\_mc}, employing a multiple-choice framework for discriminative assessment. To emulate the iterative and conversational nature of scientific inquiry, we performed synthetic data augmentation on a 20\% subset (49,663 samples) of the \texttt{Func\_ft\_train} corpus. Utilizing Llama-2-13B-Chat, we generated enriched inquiry-response pairs to form the \texttt{Func\_ft\_train\_aug} dataset, thereby enhancing the model's linguistic diversity and reasoning depth (see Appendix~\ref{apx:data_aug_methods} for methodological details). Comprehensive dataset statistics and split are detailed in Figure~\ref{fig:dataset_stats}.


\textbf{Data split.} To ensure a fair comparison with existing benchmarks, we implemented established splitting protocols for both domains.
(1) For the \texttt{Function} dataset, we adopted the partitioning strategy established by Prot2Text. Specifically, a stringent 40\% sequence identity threshold was enforced between the training and testing sets to mitigate potential data leakage and ensure rigorous evaluation of STELLA's generalization capabilities.
(2) For the \texttt{Enzyme} dataset, we followed the methodology described in IEConv~\citep{hermosilla2021intrinsic}, maintaining consistency with prior studies.

\textbf{Data explanation.} Each sample in the OPI-Struc dataset integrates a protein tertiary topology—sourced from the AFDB or PDB—with task-specific conversational instructions and their corresponding ground-truth responses. For the \texttt{Function} domain, structural data are paired with functional description derived from the UniProtKB/Swiss-Prot release 2022\_04. To rigorously assess STELLA’s generalization to newly characterized proteins, we utilized the 2024\_01 release as a temporal hold-out set for zero-shot evaluation. To mitigate option bias during multimodal training, the multiple-choice training set (\texttt{Func\_mc\_train}) was constructed by randomly permuting the four response options (A, B, C, D) for each inquiry. For evaluation, we provide two testing variants: a fixed-order version (1x) and a permuted-order version (4x). The latter employs four randomized permutations per sample to ensure a rigorous assessment of STELLA to generalize across diverse response configurations. The \texttt{Enzyme} dataset is curated from the SIFTS database~\citep{10.1093/nar/gky1114}, where original Enzyme Commission (EC) numbers are mapped to enzyme names via the BRENDA Enzyme Database(e.g., \textit{1.1.1.10} $\rightarrow$ \textit{L-xylulose reductase}). 

\textbf{Data Integrity and Contamination Mitigation.} We emphasize a critical distinction between the pre-training objective of ESM3 and our evaluation framework. While ESM3’s training involved coarse-grained protein-related keywords, OPI-Struc leverages nuanced, free-text functional narratives. This transition from keyword-level association to high-dimensional natural language descriptions ensures that the OPI-Struc test suites represent distinct evaluative scenarios. Consequently, this design effectively mitigates the risk of data contamination, ensuring that the benchmarks are not explicitly contained within ESM3’s pre-existing structural or functional priors.

\textbf{Preprocessing and Statistical Profiling.} To ensure dataset purity and reliability, OPI-Struc underwent a stringent preprocessing pipeline adhering to established data-cleaning protocols. All auxiliary metadata—including PubMed IDs, Evidence Code Ontology (ECO) IDs, and non-functional annotations—were systematically removed. We performed extensive statistical analyses to characterize the dataset's comprehensiveness. Protein sequence lengths were utilized as a proxy for structural complexity, with their distribution (illustrated in Figure~\ref{fig:function_enzyme_protein_length_density}, Appendix~\ref{apx:label_distribution_analysis}) demonstrating broad coverage across diverse folding scales. Furthermore, we analyzed the density of functional description lengths and the frequency of enzymatic labels (Figure~\ref{fig:label_distribution_analysis}, Appendix~\ref{apx:label_distribution_analysis}). These multifaceted distributions underscore the necessity for models that remain robust across varying scales of structural and functional complexity, ensuring consistent performance across the broad landscape of protein biology.

\textbf{Instruction Synthesis and Diversification.} To facilitate robust multimodal instruction tuning, the raw data were transformed into a conversational format. We sought to enhance the lexical and structural diversity of the task prompts by leveraging ChatGPT to synthesize semantically equivalent variations of the core instructions. For instance, the baseline prompt—“Please describe the function of the protein.”—was expanded into approximately 100 distinct linguistic variations. A representative list of these expanded prompts is provided in Box~\ref{box:func_instructions_1} (Appendix~\ref{detailed_instructions}). To ensure scientific accuracy and relevance, each generated variation underwent rigorous manual curation before being integrated into the final \texttt{Function} dataset. A parallel diversification process was applied to the \texttt{Enzyme} dataset (see Box~\ref{box:enzyme_instructions_1}, Appendix~\ref{detailed_instructions}). The complete diversified instructions are documented in our open-source repository.

\subsection{\textbf{Evaluation Tasks}}\label{subsec:task_def}
\textbf{Functional description prediction (\texttt{FP}).} 
This task evaluates STELLA’s capacity to decode protein tertiary topologies into detailed functional narratives. By synergistically aligning protein structural descriptors with the natural language space through MMIT, STELLA facilitates the precise generation of comprehensive biological descriptions based on 3D folding patterns. The integration of a generative LLM backbone enables dialogue-driven interactions, providing an intuitive and versatile interface for context-aware protein function elucidation.

\textbf{Enzyme-catalyzed reaction prediction (\texttt{EP}).} 
In this task, complex enzyme-catalyzed reactions are formulated as the prediction of their canonical enzyme names, which serve as semantic proxies for specific biochemical transformations mediated by the enzyme. This mapping leverages LLM's extensive knowledge of biochemical nomenclature and reaction mechanisms, ensuring that enzymatic functions are captured in a format that aligns with STELLA’s generative inference strengths. This approach bridges structured biochemical data with natural language, facilitating high-fidelity inference of catalytic roles.

\subsection{\textbf{Evaluation Metrics}}
To comprehensively evaluate STELLA's performance on the \texttt{FP} task, we employ a suite of metrics from natural language processing (NLP) to quantify the linguistic and semantic fidelity of the generated functional descriptions. Specifically, we utilize BLEU-4~\citep{papineni2002bleu} to assess $n$-gram lexical overlap between the generated and reference sequences, and ROUGE scores (1/2/L)~\citep{lin-2004-rouge} to evaluate the recall and structural preservation of biological narratives. Among these, ROUGE-L is particularly informative for functional characterization as it identifies the longest common subsequence, capturing the overarching structure of biological descriptions. To supplement these lexical-based metrics, BERTScore~\citep{zhang2019bertscore} is employed to measure token-level semantic similarity using contextual embeddings, providing a more robust assessment of functional equivalence beyond surface-level word matching.

Despite their prevalence, we recognize that standard metrics may not fully reflect the biological precision required for functional annotation. Given the current absence of universally established metrics tailored specifically for biological text generation, these metrics remain the most rigorous available proxies and have been widely adopted in prior literature. To address their inherent limitations, we further introduce MCQA for the \texttt{FP} task. This subtask utilizes Accuracy as an objective metric to evaluate STELLA’s ability to discern correct functional roles from decoys. Similarly, for the \texttt{EP} task, which involves the precise identification of catalytic roles, Accuracy is employed as the primary metric to ensure the reliability of the predicted enzyme-catalyzed reactions.

\section{\textbf{Performance Evaluation of STELLA}}
STELLA is benchmarked against five diverse scenarios—\url{FP_ft_eval}, \url{FP_ft_eval_v2401}, \url{FP_mc_eval_1x}, \url{FP_mc_eval_4x}, and \url{EP_eval}—to systematically evaluate its functional prediction capabilities. The characteristics of these test sets are summarized in Figure~\ref{fig:dataset_stats}. Comprehensive documentation of the evaluation prompts and hyperparameters can be found in Appendices~\ref{apx:prompt_template_testing} and \ref{apx:hyperpara_train_test}.

\subsection{\textbf{Evaluation of FP Performance}}

\begin{figure}[!htbp]
    \vspace{-0.1in}
    \centering
    \includegraphics[width=0.9\linewidth]{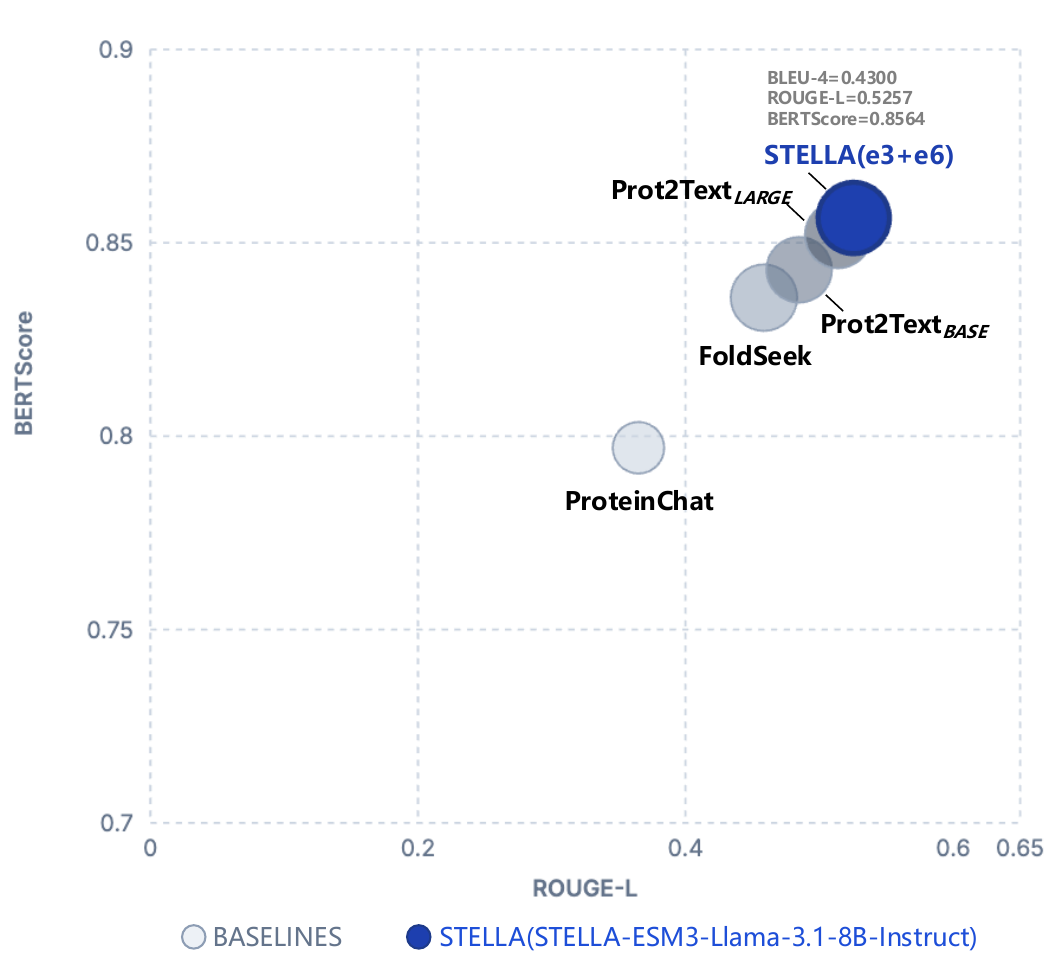}
    \caption{\textbf{Hold-out evaluation of FP performance across STELLA, baselines and state-of-the-art methods on the \texttt{FP\_ft\_eval} benchmark.} STELLA is fine-tuned on the \texttt{Func\_ft\_train} dataset using a two-stage strategy (3+6 epochs).}
    \label{fig:results_FP_FTQA_eval}
    \vspace{-0.1in}
\end{figure}

\textbf{Hold-out Benchmark.} To establish a baseline of STELLA’s performance, we first evaluated the model on the independent hold-out test set, \texttt{Func\_ft\_test}, following the experimental protocol established by Prot2Text. As summarized in Figure~\ref{fig:results_FP_FTQA_eval} (see Table~\ref{apx:tab:results_hold_out_eval}, Appendix \ref{apx:results_hold_out_eval} for detailed results), STELLA (e3+e6) achieves state-of-the-art results, consistently outperforming prior baselines.

\begin{figure}
    \centering
    \includegraphics[width=0.9\linewidth]{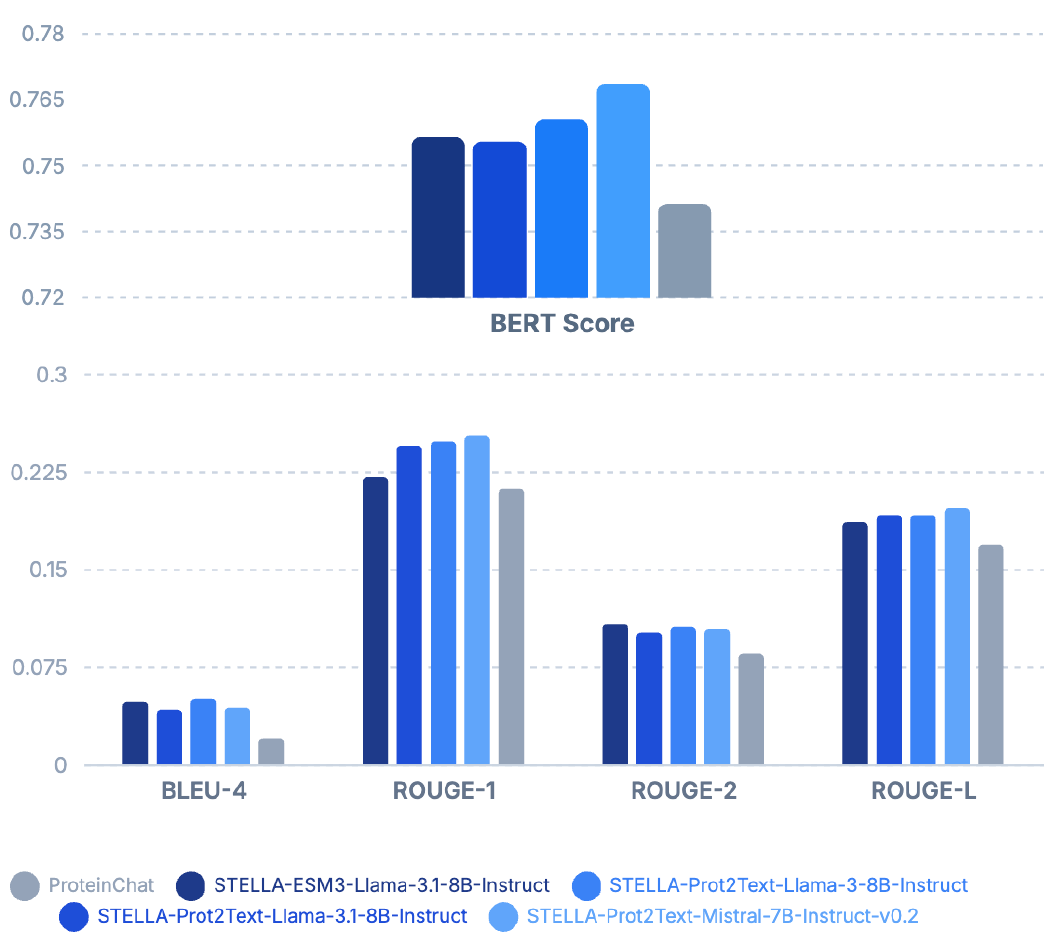}
    \caption{\textbf{Zero-shot temporal evaluation of STELLA on the \texttt{FP\_ft\_eval\_v2401} benchmark.} STELLA models are fine-tuned on the \texttt{Func\_ft\_train} dataset using a two-stage strategy (3+3 epochs).}
    \label{fig:zero_shot_eval}
    \vspace{-0.2in}
\end{figure}

\textbf{Zero-shot Temporal Generalization.} We further assessed STELLA’s robustness against distributional shifts using \texttt{FP\_ft\_test\_v2401}, a temporal hold-out set derived from a newer Swiss-Prot release unseen during training. As shown in Figure~\ref{fig:zero_shot_eval}, we observed a performance decrement of zero-shot performance relative to the hold-out evaluation (see Table~\ref{apx:tab:results_zero_shot_eval}, Appendix \ref{apx:results_zero_shot_eval} for more detailed results). This is likely attributable to newly characterized proteins possessing novel structural or functional motifs that remain underrepresented in the training corpus—a common challenge in protein representation learning due to the continuous expansion of biological knowledge. To address this Out-of-Distribution (OOD) gap, future iterations will explore RAG strategies and the integration of external functional metadata.

\textbf{Robustness to Structural Degradation}. Inherent noise and incomplete coordinates are prevalent in experimental protein structures. To evaluate STELLA’s resilience, we simulated structural perturbations by truncating the terminal 10\% of residues for each protein in the test set. As detailed in Table~\ref{tab:robustness_incomplete}, STELLA demonstrates remarkable stability under these degraded conditions. Notably, for the e3+e6 configuration, the ROUGE-L score decreased by only 4.1\% (from 0.5257 to 0.4915), while Prot2Text$_{LARGE}$ experienced a more substantial decline of 13.7\%. Given that all models were trained on complete structures, STELLA’s superior retention of functional information highlights its representational robustness and applicability to partially resolved structural data.

\begin{table}[!htbp]
    \centering
    \caption{\textbf{Performance evaluation on structural degradation.} ROUGE-L are reported for each model on complete and incomplete protein. STELLA (STELLA-ESM3-Llama-3.1-8B-Instruct) is fine-tuned on the \texttt{Func\_ft\_train} dataset using a two-stage strategy.}
    \label{tab:robustness_incomplete}
    \vspace{-0.1in}
    \renewcommand{\arraystretch}{0.85} 
    \setlength{\tabcolsep}{5pt} 
    \resizebox{\linewidth}{!}{
    \small
    \begin{tabular}{lccc}
    \toprule
    Model & \makecell{Complete} & \makecell{Incomplete} & \makecell{Perf. Drop} \\
    \midrule
    Prot2Text$_{LARGE}$ & 0.5140 & 0.4438 & 13.7\% \\
    STELLA (e3+e3) & 0.5041 & 0.4805 & 4.7\% \\
    STELLA (e3+e6) & 0.5257 & 0.4915 & 4.1\% \\
    \bottomrule
    \end{tabular}
    }
    \vspace{-0.1in}
\end{table}

\textbf{Comparison with Alignment-based Retrieval.} We compared STELLA against a representative retrieval-based baseline utilizing Foldseek~\citep{van2024fast}, which includes two steps: structure retrieval using Foldseek and function mapping from Swiss-Prot. This pipeline involves (1) structural retrieval via Foldseek (easy-search mode, $e\text{-}value < 0.001$) within the training set, followed by (2) functional mapping from the top-1 retrieved Swiss-Prot hit. The median $e\text{-}value$ for the top hits was $2.723 \times 10^{20}$, reflecting high retrieval confidence. Nevertheless, as shown in Figure~\ref{fig:results_FP_FTQA_eval}, STELLA outperforms the Foldseek-based method (by 14.6\%) in ROUGE-L. This underscores STELLA’s ability to generalize beyond simple homology-based lookups by learning the underlying sequence-structure-function mapping.

\textbf{Objective Assessment via MCQA.} To mitigate the impact of linguistic variability—where semantically correct responses may diverge from reference texts in phrasing—we introduced the MCQA subtask. This format eliminates generative ambiguity and provides a standardized metric for discriminative reasoning. STELLA achieved accuracies of \textbf{80.56} and \textbf{76.18} on \texttt{FP\_mc\_eval\_1x} and \texttt{FP\_mc\_eval\_4x}, respectively. The results demonstrate STELLA’s robust instruction-following capability and its capacity to perform precise functional reasoning under constrained choice sets.

\textbf{Generalization Analysis Across Structural Homology Tiers.} To assess structural generalization beyond sequence similarity, we quantified the representational density of structural motifs via Foldseek-based clustering. As detailed in Table \ref{apx:tab:struc_density}, Appendix \ref{apx:struc_density}, STELLA maintains robust performance even for proteins with minimal structural homology in the training set.

\subsection{\textbf{Evaluation of EP Performance}}
The \texttt{EP\_eval} suite was employed to determine STELLA’s precision in predicting enzyme-catalyzed reactions. During preprocessing, 10 samples were excluded due to missing atomic coordinates, ensuring the integrity of the structural embeddings. As illustrated in Table~\ref{tab:results_EP_eval}, extending the Stage 2 training duration from 3 to 6 epochs yielded a performance gain, increasing accuracy from \textbf{88.06} to \textbf{88.85}. Ultimately, STELLA establishes a new state-of-the-art for the \texttt{EP} task, surpassing leading models such as CDConv~\citep{fan2022continuous} and Sable~\citep{10.1093/bib/bbaf120} (previous best: 88.50). This performance highlights STELLA’s exceptional capability in bridging high-resolution structural features with precise biochemical nomenclature.

\begin{table}[!htbp]
  \caption{\textbf{Evaluation of EP performance.} Accuracy measures exact matches with the ground truth. STELLA (STELLA-ESM3-Llama-3.1-8B-Instruct) is fine-tuned on the \texttt{Enzyme\_train} dataset using a two-stage strategy. \textbf{Bold}: best; \underline{underline}: runner-up.}
  \vspace{-0.1in}
  \label{tab:results_EP_eval}
  \centering
  \renewcommand{\arraystretch}{0.85} 
  \setlength{\tabcolsep}{5pt} 
  \resizebox{\linewidth}{!}{
  \begin{tabular}{lc} 
    \toprule
    Model & Accuracy $ \uparrow $  \\
    \midrule
    \multicolumn{2}{l}{\textit{w/o pretrain}} \\
    \midrule
    UniRep~\citep{alley2019unified} & 72.90  \\
    3DCNN~\citep{derevyanko2018deep} & 78.80  \\
    TAPE-LSTM~\citep{NEURIPS2019_37f65c06} & 79.90  \\
    HH-suite3~\citep{steinegger2019hh} & 82.60  \\
    GearNet-Edge-IEConv~\citep{zhang2022protein} & 85.30 \\
    IEConv~\citep{hermosilla2021intrinsic} & 87.20  \\
    New IEConv~\citep{Hermosilla2022ContrastiveRL} & 87.20  \\
    CDConv~\citep{fan2022continuous} & \underline{88.50} \\
    \midrule
    \multicolumn{2}{l}{\textit{w/ pretrain}} \\
    \midrule
    DeepFRI~\citep{gligorijevic2021structure} & 63.30 \\
    ProtBERT-BFD~\citep{9477085} & 72.20 \\
    ESM-1b~\citep{doi:10.1073/pnas.2016239118} & 83.10 \\
    GearNet-Multiview-Contrast~\citep{zhang2022protein} & 87.50  \\
    New IEConv~\citep{hermosilla2022contrastive} & 88.10  \\
    Sable~\cite{10.1093/bib/bbaf120} & \underline{88.50}  \\
    \midrule
    \multicolumn{2}{l}{\textit{MMIT (Two-stage training on \texttt{Enzyme\_train} dataset)}} \\
    \midrule
    STELLA (e3+e3) & 88.06  \\
    STELLA (e3+e6) & \textbf{88.85} \\
    \bottomrule
  \end{tabular}
  }
  \vspace{-0.3in}
\end{table}

\section{\textbf{Comparative Analysis}}  
\subsection{\textbf{Encoder Efficacy and LLM Synergy}} 
We characterized the discriminative power of protein encoders by visualizing the embeddings of the \texttt{Func\_ft\_test} set via UMAP. Compared to Prot2Text and SaProt, ESM3 produces more discrete and compact functional clusters (Figure~\ref{fig:umap_function_test}), reflecting a more biologically informative latent space. A comparative study of various LLMs within the STELLA framework reveals that the integration of Llama-3.1-8B-Instruct yields superior performance in \texttt{FP} and \texttt{EP} tasks (Table~\ref{tab:ablation_prot_enc_and_llms}).

Additionally, our analysis of reasoning-centric LLMs and standard models indicates that while reasoning capabilities slightly bolster OOD robustness, the overall performance in these challenging regimes remains a frontier (Table~\ref{tab:compare_reason_and_non_reason_llms}). These findings are consistent with recent studies on the stability and generalization of reasoning models in specialized domains~\citep{yao2025unveiling, wang2025beyond, huang2025thinkbench}, highlighting the ongoing challenge of achieving robust zero-shot reasoning in protein biology.

\begin{table*}[!htbp]
  \caption{\textbf{Comparison of protein encoder efficacy and LLM synergy on the \texttt{FP\_ft\_eval} benchmark.} All models are fine-tuned on \texttt{Func\_ft\_train} using a two-stage strategy (3+3 epochs). \textbf{Bold}: best; \underline{underline}: runner-up.}
  \vspace{-0.1in}
  \label{tab:ablation_prot_enc_and_llms}
  \centering
  \small
  \renewcommand{\arraystretch}{0.85} 
  \setlength{\tabcolsep}{5pt} 
  \resizebox{\linewidth}{!}{
  \begin{tabular}{lccccc}
    \toprule
    \multirow{2}{*}{Model} 
    & \multirow{2}{*}{BLEU-4 $ \uparrow $} 
    & \multirow{2}{*}{BERTScore $ \uparrow $} 
    & \multicolumn{3}{c}{ROUGE Score $ \uparrow $} \\  
    \cmidrule(r){4-6}
    & & & ROUGE-1 & ROUGE-2 & ROUGE-L \\ 
    \toprule
    \multicolumn{6}{l}{\textbf{ESM3 encoder}} \\
    \midrule
    STELLA-ESM3-Llama-3.1-8B-Instruct & \textbf{0.4024} & 0.8496 & 0.5218 & \textbf{0.4487} & \textbf{0.5041} \\
    STELLA-ESM3-Llama-3-8B-Instruct   & \underline{0.4020} & 0.8503 & 0.5138 & \underline{0.4478} & 0.5001 \\
    STELLA-ESM3-Phi-3-mini-128k-instruct & 0.3807 & 0.8435 & 0.4991 & 0.4273 & 0.4839 \\

    \midrule
    \multicolumn{6}{l}{\textbf{Prot2Text encoder}} \\
    \midrule
    STELLA-Prot2Text-Llama-3.1-8B-Instruct & 0.4009 & 0.8497 & \textbf{0.5284} & 0.4454 & \underline{0.5031} \\
    STELLA-Prot2Text-Llama-3-8B-Instruct   & 0.3892 & 0.8456 & 0.5177 & 0.4329 & 0.4915 \\
    STELLA-Prot2Text-Phi-3-mini-128k-instruct & 0.3771 & 0.8426 & 0.5058 & 0.4210 & 0.4799 \\
    STELLA-Prot2Text-Mistral-7B-Instruct-v0.2 & 0.3889 & \underline{0.8525} & 0.5224 & 0.4359 & 0.4949 \\
    STELLA-Prot2Text-BioMedGPT-LM-7B & 0.3999 & 0.8488 & \underline{0.5282} & 0.4447 & 0.5020 \\
    STELLA-Prot2Text-BioMistral-7B-DARE & 0.3870 & \textbf{0.8533} & 0.5241 & 0.4357 & 0.4980 \\

    \midrule
    \multicolumn{6}{l}{\textbf{SaProt encoder}} \\
    \midrule
    STELLA-SaProt-Llama-3-8B-Instruct & 0.3588 & 0.8276 & 0.4685 & 0.3965 & 0.4523 \\
    STELLA-SaProt-Mistral-7B-Instruct-v0.2 & 0.3514 & 0.8251 & 0.4607 & 0.3894 & 0.4455 \\

    \bottomrule
  \end{tabular}
  }
  \vspace{-0.2in}
\end{table*}

\begin{table}[!htbp]
  \vspace{-0.3in}
  \caption{\textbf{Comparison of Reasoning-centric vs. Standard LLMs.} All models utilize the STELLA-ESM3 architecture and are fine-tuned on the \texttt{Func\_ft\_train} dataset using a two-stage strategy (3+6 epochs). \textbf{Bold}: best; \underline{underline}: runner-up. DS: DeepSeek.}
  \label{tab:compare_reason_and_non_reason_llms}
  \centering
  \renewcommand{\arraystretch}{0.85} 
  \setlength{\tabcolsep}{5pt} 
  \resizebox{\linewidth}{!}{
  \begin{tabular}{lcccc} 
    \toprule
    LLM & BLEU-4 $ \uparrow $ & BERTScore $ \uparrow $ & ROUGE-L $ \uparrow $ \\  
    \midrule
    \multicolumn{4}{l}{\textit{Evaluation on \texttt{FP\_ft\_eval} (Hold-out)}} \\
    \midrule
    Llama-3.1-8B-Instruct & \textbf{0.4300} & \underline{0.8564} & \textbf{0.5257} \\
    DS-R1-Distill-Qwen-1.5B & 0.3869 & 0.8422 & 0.4853 \\
    DS-R1-Distill-Qwen-14B & \underline{0.4268} & 0.8549 & 0.5215 \\
    DS-R1-Distill-Llama-8B & 0.4249 & \textbf{0.8569} & \underline{0.5229} \\
    \midrule
    \multicolumn{4}{l}{\textit{Evaluation on \texttt{FP\_ft\_eval\_v2401} (OOD)}} \\
    \midrule
    Llama-3.1-8B-Instruct & 0.0489 & 0.7565 & \underline{0.1867} \\
    DS-R1-Distill-Qwen-1.5B & 0.0468 & 0.7566 & 0.1774 \\
    DS-R1-Distill-Qwen-14B & \underline{0.0538} & \underline{0.7588} & 0.1845 \\
    DS-R1-Distill-Llama-8B & \textbf{0.0542} & \textbf{0.7593} & \textbf{0.1879} \\
    \bottomrule
  \end{tabular}
  }
  \vspace{-0.2in}
\end{table}

\subsection{\textbf{Stage-wise Training Strategy}} 

The training procedure for STELLA is partitioned into two distinct phases: cross-modal representation alignment and multimodal instruction tuning. This decoupled strategy is designed to mitigate optimization conflicts and ensure a stabilized transition from protein-centric descriptors to complex natural language reasoning.

\begin{figure}[!htbp]
    \vspace{-0.2in}
    \centering
    \includegraphics[width=0.9\linewidth]{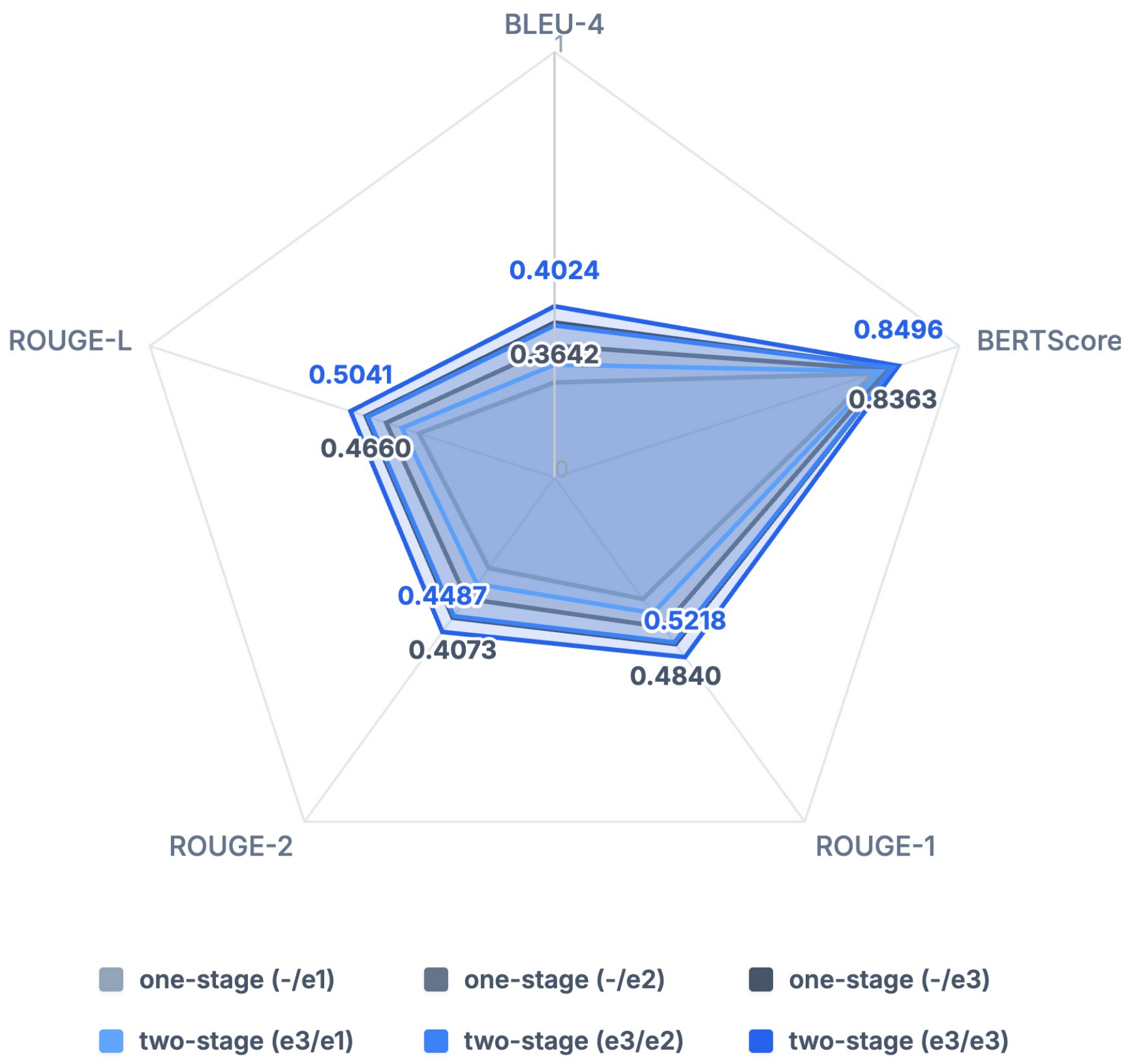}
    \caption{\textbf{Comparative analysis of stage-wise training strategy for STELLA-ESM3-Llama-3.1-8B-Instruct on \texttt{FP\_ft\_eval}}. The two-stage strategy consistently outperforms the single-stage training approach.}
    \label{fig:ablation_training_stages}
    \vspace{-0.2in}
\end{figure}
\textbf{Stage 1} focuses on projecting bimodal protein embeddings into the LLM’s latent space via a modality connector. By aligning protein features with textual semantics at this stage, we bridge the inherent cross-modal disparity, allowing the LLM to interpret biological features as coherent linguistic tokens. \textbf{Stage 2} emphasizes instruction tuning and task-specific refinement, enhancing the model’s generative fidelity and zero-shot generalization. This progressive paradigm is instrumental in preventing catastrophic forgetting and representation collapse; specifically, it ensures that the model does not disproportionately overfit to textual priors at the expense of intrinsic protein features. Furthermore, Stage 2 accommodates diverse response schemata, such as MCQA templates, to facilitate more controlled and task-aligned outputs. While both stages utilize the same dataset, they employ differential learning rates and parameter freezing strategies to facilitate optimal convergence. As evidenced in Figure~\ref{fig:ablation_training_stages}, this two-stage paradigm consistently yields superior performance compared to a single-stage joint-training approach across all evaluated metrics for the \texttt{FP} task (see Table~\ref{apx:tab:ablation_training_stages}, Appendix \ref{apx:ablation_training_stages} for more detailed results).

    
    
      

\section{\textbf{Conclusion and Future Work}}

This study introduces STELLA, a multimodal LLM tailored for protein function elucidation, supported by the curated OPI-Struc dataset—a high-fidelity resource specifically designed for multimodal instruction tuning. By synergistically aligning bimodal sequence-structure representations with the extensive world knowledge embedded in LLMs, STELLA not only streamlines the multimodal training pipeline but also defines a new state-of-the-art benchmark in two critical domains: functional description and enzyme-catalyzed reaction prediction. Beyond its empirical performance, STELLA represents a significant paradigm shift in computational protein science~\citep{fan2025computational}, effectively bridging the chasm between generative AI and life sciences. Our findings demonstrate that the integration of structural topologies and natural language reasoning can transcend the limitations of traditional protein language models. Looking ahead, we envision that the development of domain-specific biomolecular tokenizers and the deployment of autonomous agentic AI will unlock unprecedented analytical capabilities in biomedical discovery.

\section*{Limitations}
Despite its advancements, STELLA is not without limitations. Currently, the model's performance is bounded by the granularity of the structural \textbf{tokenization} and the inherent complexity of aligning high-dimensional geometric features with discrete linguistic tokens. While effective, the current framework primarily relies on general-purpose LLM reasoning, which may occasionally lack the hyper-specialized biochemical intuition required for \textit{de novo} functional discovery.
Furthermore, the current iteration does not yet incorporate \textbf{external knowledge integration}, which could provide real-time access to the rapidly expanding body of biological literature and structural databases. Future research will focus on developing high-resolution structural adapters, exploring RAG-enhanced workflows to mitigate hallucinations, and investigating multi-agent architectures to handle complex, multi-step biological reasoning tasks. In addtion, well-established \textbf{benchmark} suite remains limited. The aforementioned refinements will further solidify multimodal LLMs and agentic AI as indispensable tools for accelerating innovation within the vast and intricate landscape of life sciences.

\bibliography{bibliography}

\appendix
\section*{\textbf{Appendix}}

\section{Prompt Templates for Training}\label{apx:prompt_template_training}

\begin{mybox}[label={box:prompt_llama3.1}, top=1mm, bottom=1mm, left=1mm, right=1mm]{The prompt template for STELLA-ESM3-Llama-3.1-8B-Instruct}
\begin{lstlisting}
<|begin\_of\_text|><|start\_header\_id|>user<|end\_header\_id|>

<structure>
May I request a comprehensive breakdown outlining the function linked to the protein? 

<|eot\_id|><|start\_header\_id|>assistant<|end\_header\_id|>

Involved in the gluconeogenesis. Catalyzes stereospecifically the conversion of dihydroxyacetone phosphate (DHAP) to D-glyceraldehyde-3-phosphate (G3P). <|eot\_id|><|end\_of\_text|>
\end{lstlisting}
\end{mybox}

\begin{mybox}[label={box:prompt_mistral},top=1mm, bottom=1mm, left=1mm, right=1mm]{The prompt template for STELLA-Prot2Text-Mistral-7B-Instruct-v0.2}
\begin{lstlisting}
<s>[INST] <structure>
May I request a comprehensive breakdown outlining the function linked to the protein? [/INST]Involved in the gluconeogenesis. Catalyzes stereospecifically the conversion of dihydroxyacetone phosphate (DHAP) to D-glyceraldehyde-3-phosphate (G3P)</s>
\end{lstlisting}
\end{mybox}

\section{Prompts for Evaluation}\label{apx:prompt_template_testing}

We design the following evaluation prompts to constrain the model output in specific tasks.  

\begin{mybox}[label={box:prompt_fp_ft}, top=1mm, bottom=1mm, left=1mm, right=1mm]{Evaluation prompt for \texttt{FP\_ft}}
\begin{lstlisting}
<user>
What are the main functions of this protein?
\end{lstlisting}
\end{mybox}

\begin{mybox}[label={box:prompt_fp_mc}, top=1mm, bottom=1mm, left=1mm, right=1mm]{Evaluation prompt for \texttt{FP\_mc}}
\begin{lstlisting}
<user>
Please answer the question directly with the corresponding letter (A, B, C, or D) from the provided options.
\end{lstlisting}
\end{mybox}

\begin{mybox}[label={box:prompt_ep}, top=1mm, bottom=1mm, left=1mm, right=1mm]{Evaluation prompt for \texttt{EP}}
\begin{lstlisting}
<user>
What is the enzyme name linked to this protein?
\end{lstlisting}
\end{mybox}

\vspace{-0.5cm}
\section{Analysis of Data Label Distribution of OPI-Struc}\label{apx:label_distribution_analysis}

\begin{figure*}[!htbp]
    \centering
    \includegraphics[width=0.9\linewidth]{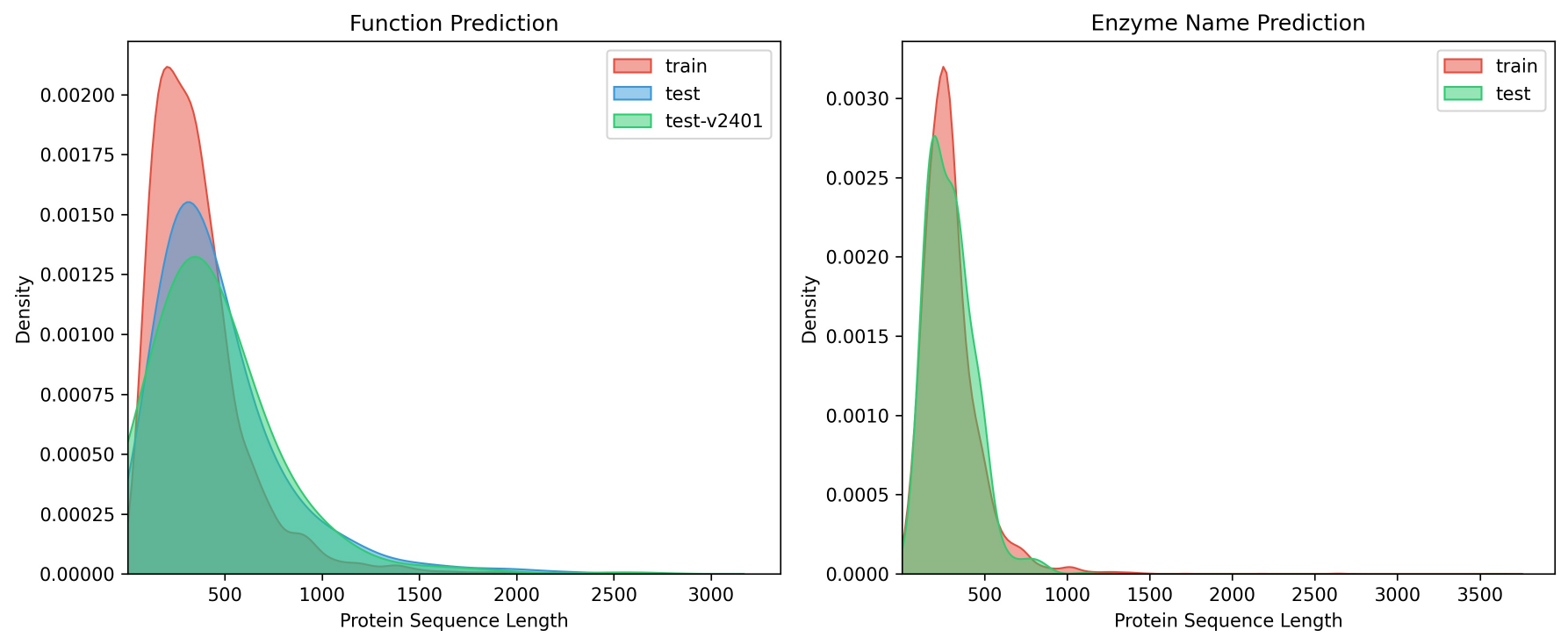}
    \vspace{-0.2in}
    \caption{\textbf{Distribution of protein sequence lengths across the \texttt{FP} (left) and \texttt{EP} (right) tasks for training and testing sets.} The variation in sequence length distribution between the training and testing sets ensures model robustness across proteins with diverse structural complexities.}
    \label{fig:function_enzyme_protein_length_density}
\vspace{-0.1in}
\end{figure*}

\begin{figure}[!htbp]
    \centering
    \vspace{-0.5cm}
    \begin{subfigure}{0.43\textwidth}
        \includegraphics[width=0.9\linewidth]{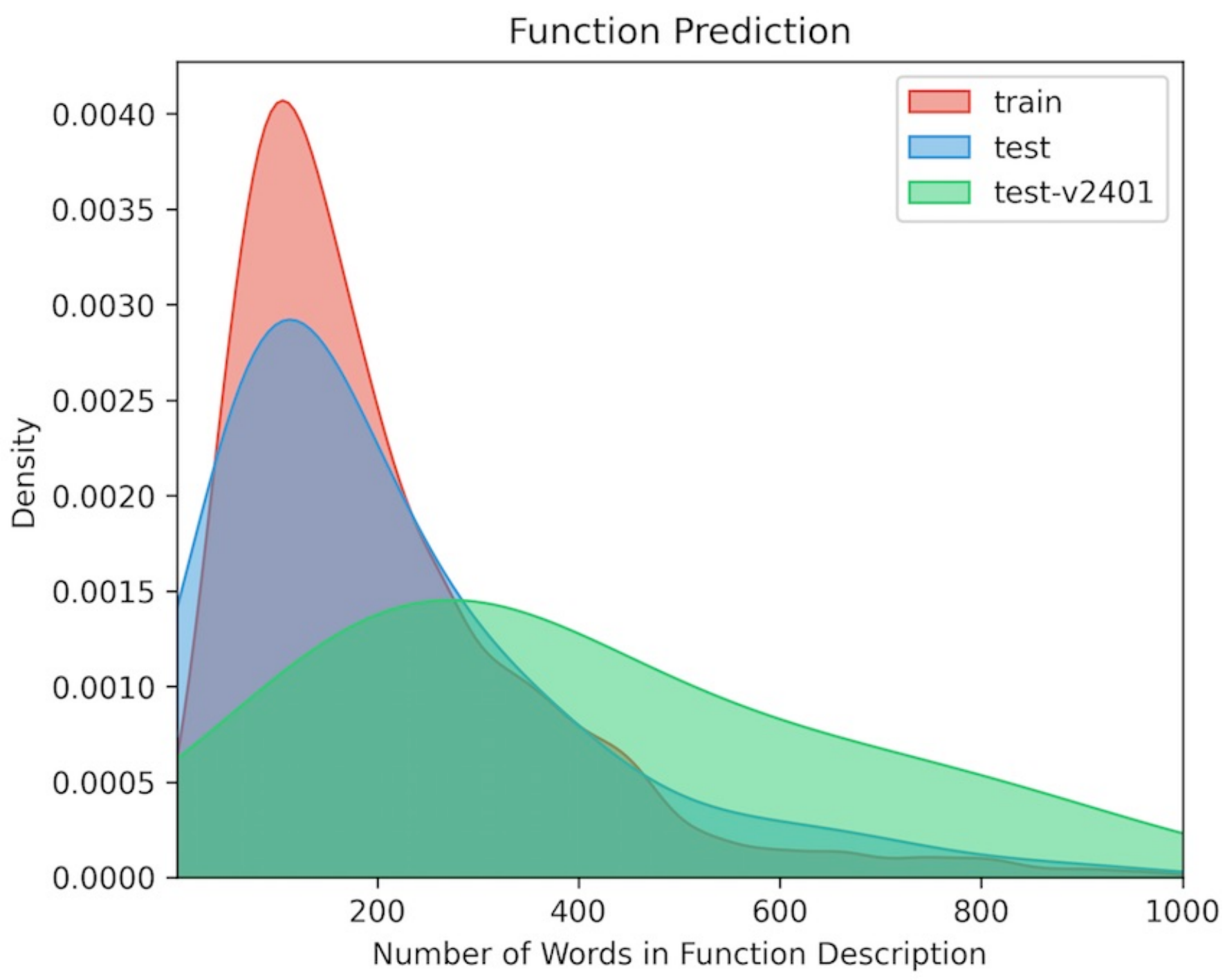}
        \caption{Length distribution of functional descriptions in the \texttt{Function} dataset.}
        \label{fig:function_description_length_density}
    \end{subfigure}
    \hspace{0.5cm}
    \begin{subfigure}{0.4\textwidth}
        \includegraphics[width=0.9\linewidth]{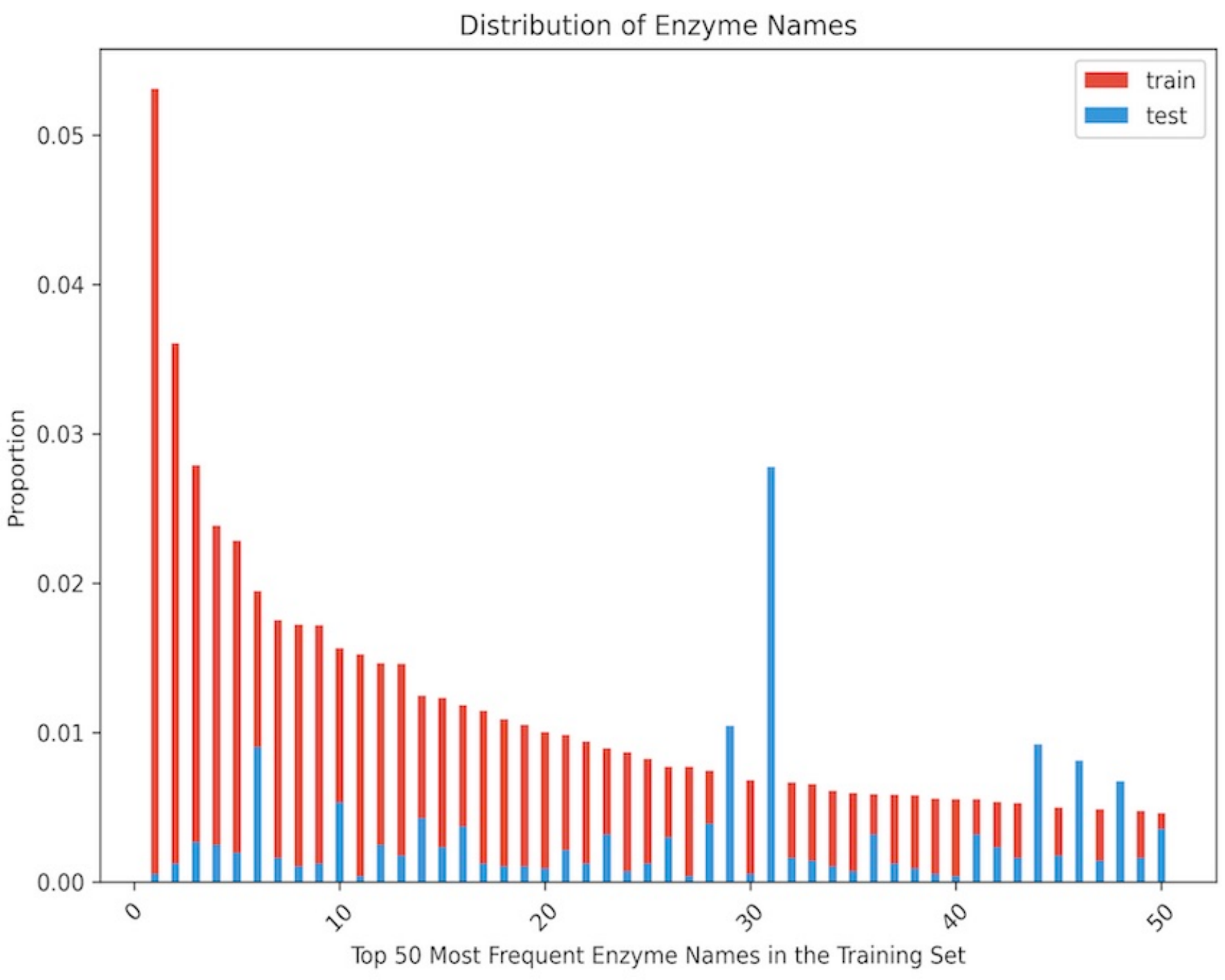}
        \caption{Frequency of enzyme names in the \texttt{Enzyme} dataset. The training set follows a long-tailed pattern, but the test set distribution differs significantly.}
        \label{fig:enzyme_name_distribution}
    \end{subfigure}
    \caption{Analysis of label distributions across datasets.}
    \label{fig:label_distribution_analysis}
\end{figure}

Figures~\ref{fig:function_enzyme_protein_length_density} illustrates the distribution of protein sequence lengths across the \texttt{FP} (left) and \texttt{EP} (right) tasks for training and testing sets. Figure~\ref{fig:label_distribution_analysis} shows (a) the length distribution of functional descriptions in the Function dataset and (b) the frequency of enzyme names in the Enzyme dataset. 

\vspace{-0.2cm}
\section{Performance Sensitivity to Structural Representational Density}\label{apx:struc_density}
To delineate the relationship between structural familiarity and predictive accuracy, we performed a clustering-based sensitivity analysis. We first partitioned the training set into structural clusters using Foldseek based on global fold similarity. For each protein in the test suite (\texttt{Func\_ft\_test}, N=4,203), we identified its corresponding structural cluster within the training corpus. The \textbf{representational density}—defined as the number of structurally homologous training samples within the matched cluster—serves as a proxy for structural novelty; a smaller cluster size indicates a more novel topology relative to the training distribution.

As summarized in Table \ref{apx:tab:struc_density}, we observed a positive correlation between representational density and predictive performance. Specifically, the mean ROUGE-L score increases from 0.4323 for near-unique structures (cluster size $\leq$ 1) to 0.6691 for well-represented protein families ($\ge$ 20 samples).  Notably, STELLA retains substantial generative capacity even in the most challenging "structural dark matter" regime (0-1 similar training samples), underscoring its ability to generalize beyond simple structural memorization to intrinsic sequence-structure-function mappings.

This performance gradient suggests that while STELLA benefits from structural motifs encountered during training, its multimodal alignment enables the inference of biological roles for proteins with unseen global folds.

\begin{table}[!ht]
    \centering
    \caption{Performance stratification by structural representational density. Test samples are grouped based on the number of structural homologs identified in the training set via Foldseek.}
    \label{apx:tab:struc_density}
    \resizebox{\linewidth}{!}{
    \begin{tabular}{ccc}
       \toprule
       \makecell[c]{Representational Density \\ (Train Samples per Cluster)} & 
       \makecell[c]{Testing \\ samples} & 
       \makecell[c]{Mean \\ ROUGE-L} \\
       \midrule
       $(0, 1]$      & 1202 & 0.4323 \\
       $[2, 5]$      & 1205 & 0.4918 \\
       $[6, 19]$     & 921  & 0.5558 \\
       $[20, 1134]$  & 875  & 0.6691 \\
       \bottomrule
    \end{tabular}
    }
    \vspace{-0.5cm}
\end{table}

\section{Hyperparameters for Training and Evaluation}\label{apx:hyperpara_train_test} 

The stage1 training aims to align the embedding space of protein structures and texts. In this stage, the modality connector is trainable, while both the protein structure encoder and the LLM are frozen. Stage2 is dedicated to enabling STELLA to follow complicated natural language instructions and generate response dedicated to protein tasks. In this stage, both the modality connector and the LLM are trained with different learning rates, while the protein structure encoder is still frozen. Both stages use the same training datasets. The training prompt templates follow the examples shown in Appendix~\ref{apx:prompt_template_training}.

The hyperparameters in two stages are summarized in Table~\ref{tab:hyperpara_train_test}. It should be noted that we adopt different learning rates for each different components of STELLA to finely control the training process. Especially, in stage2, we set the learning rate of the modality connector larger than LLM backbone, to improve LLMs' training convergence.
\begin{table*}[!ht]
    \small
    \centering
    \caption{\textbf{Hyperparameters for stage1 training, stage2 training and testing.} FFT: Full Fine-tuning.}
    \label{tab:hyperpara_train_test}
    \begin{tabularx}{\textwidth}{Xccc}
       \toprule
       \textbf{Config} & \textbf{Stage1} & \textbf{Stage2} & \textbf{Testing} \\
       \midrule
       DeepSpeed\ ZeRO Stage  & 2 & 3 & N/A \\
       optimizer & AdamW & AdamW & N/A \\
       optimizer hyperparameters & ($\beta_1$,$\beta_2$)=(0.9, 0.999), eps=1e-8 & ($\beta_1$,$\beta_2$)=(0.9, 0.999), eps=1e-8  & N/A \\
       per\_device\_train\_batch\_size & 2 & 1(FFT)/2(LoRA) & N/A \\
       gradient\_accumulation\_steps & 4 & 2(FFT)/4(LoRA) & N/A \\
       gradient\_checkpointing & True & True & N/A \\
       learning rate (lr)  & 2e-5 (Connector) & 2e-4 (Connector), 2e-5 (LLM) & N/A \\
       weight\ decay  & 0.0 & 0.0 & N/A \\
       warmup\ steps  & 48 & - & N/A \\
       warmup\ ratio  & - & 0.03 & N/A \\
       lr scheduler type  & cosine & cosine & N/A \\
       training\ epochs & 3 & 3 & N/A \\
        GPU & 4*A100 & 8*A100(FFT)/4*A100(LoRA) & 1*A100 \\
       temperature & N/A & N/A & 0.2 \\
       top\_k & N/A & N/A & 50 \\
       top\_p & N/A & N/A & 0.75 \\
       num\_beams & N/A & N/A  & 1 \\
       max\_new\_tokens & N/A & N/A & 1000 \\
       use\_cache & N/A & N/A & True \\
       do\_sample & N/A & N/A & True \\
       \bottomrule
    \end{tabularx}
    \vspace{-0.5cm}
\end{table*}

\vspace{-0.5cm}
\section{Comparison of Protein Encoders}\label{apx:prot_encoder_compare}

In terms of STELLA's architecture, we employ three protein encoders--ESM3~\citep{hayes2024simulating}, Prot2Text~\citep{abdine2023prot2text}, and SaProt~\citep{su2023saprot}--for comparative analysis. ESM3 and Prot2Text model the interplay of sequence, structure, and function, while SaProt only models the sequence and structure modalities. This setup allows us to investigate the impact of different encoders on the STELLA's overall performance, providing insights into the contributions of different components to the its capability.

\textbf{ESM3} is a large multimodal model pretrained on massive sequence, structure, and function tokens using masked language modeling (MLM) strategy. It encodes these modalities as discrete token tracks, which are fused into a unified representation space via several transformer blocks, with geometric attention in the first block to incorporate atomic information.

\textbf{Prot2Text} is a multimodal model that integrates a Relational Graph Convolution Network (RGCN), ESM-2, and GPT-2 to generate protein function annotation. It combines two sources of information: the output of the RGCN, which processes all-atom protein structures to provide detailed structural representations, and protein sequences processed by ESM-2. The Prot2Text encoder aligns these integrated data with functional annotation through a generative alignment approach using a text decoder. 

\textbf{SaProt} is a large-scale pre-trained pLM utilizing around 40 million protein sequences and structures, with a structure-aware vocabulary integrating residue tokens and structural tokens simultaneously. It adopts an ESM-based architecture that takes structure-aware protein sequences as input, which combine protein sequence residue tokens and discrete structural tokens encoded via FoldSeek~\citep{van2024fast}. However, this encoder is not aligned with functional annotation text.

\begin{figure*}[!ht]
    \centering
    \includegraphics[width=\textwidth]{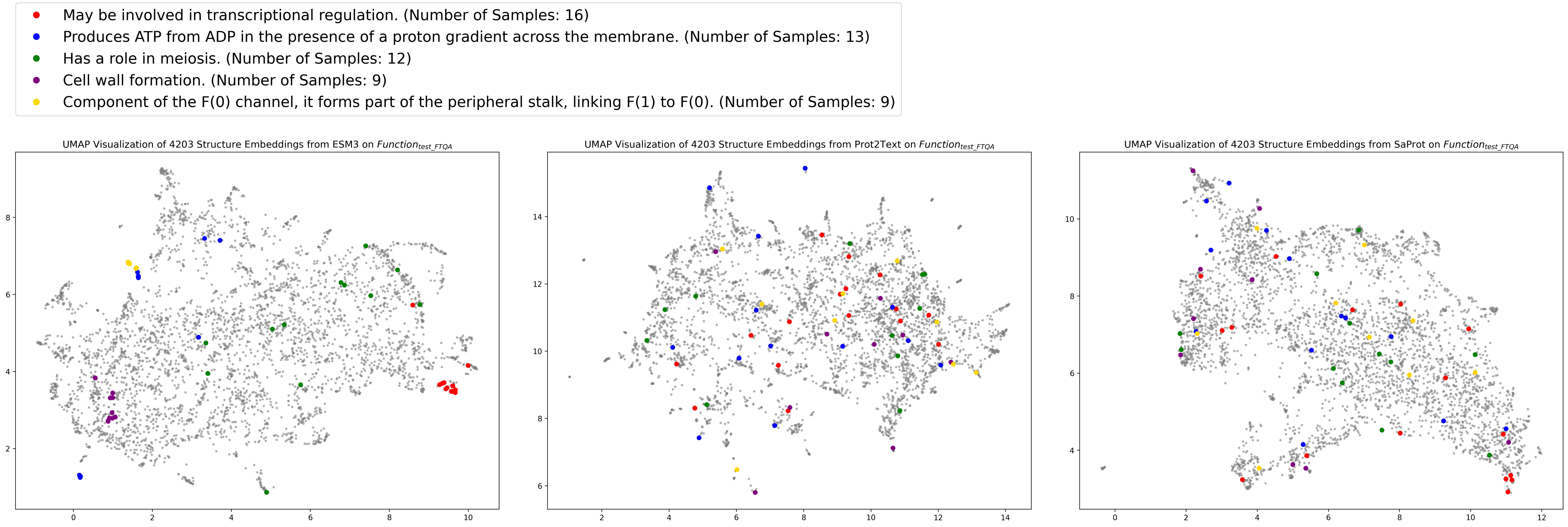}
    \vspace{-0.15in}
    \caption{\textbf{UMAP visualization of 4,203 protein structure embeddings in the testing set \texttt{Func\_ft\_test} generated by ESM3, Prot2Text, and SaProt.} Each plot illustrates the clustering of protein structures based on their embeddings, revealing the representational differences among the three encoders. The highlighted proteins belong to specific functions as detailed in the legend. ESM3 demonstrates the strongest representative ability.}
    \label{fig:umap_function_test}
\end{figure*}

Figure~\ref{fig:umap_function_test} illustrates the UMAP visualization of protein embeddings generated by ESM3, Prot2Text, and SaProt for the 4,203 testing samples in \texttt{Func\_ft\_test}, from which it can be observed that ESM3 provides more distinct and informative protein feature representations.

\section{Different Composition of Protein Encoders and LLMs}\label{apx:composition_prot_enc_llm}
The architecture of STELLA is flexible and customizable to integrate various protein encoders and LLMs to form variants. We elaborately choose different protein encoders and LLMs to investigate the effectiveness of different composition of these components, as shown in Table \ref{tab:composition_prot_enc_llm}.

\begin{table*}[!ht]
  \caption{\textbf{Specifications of STELLA composition of various protein structure encoders and foundation LLMs.}}
  \label{tab:composition_prot_enc_llm}
  \centering
  \resizebox{\linewidth}{!}{
  \begin{threeparttable}
  \begin{tabular}{llll} 
    \toprule
    Protein encoder & LLM & Note for LLM & STELLA variant \\

    \midrule
    \multirow{5}{*}{ESM3}
    & Llama-3.1-8B-Instruct~\citep{llama3modelcard} & Open source by Meta & STELLA-ESM3-Llama-3.1-8B-Instruct \\
    & Llama-3-8B-Instruct~\citep{llama3modelcard} & Open source by Meta & STELLA-ESM3-Llama-3-8B-Instruct \\
    & Mistral-7B-Instruct-v0.2~\citep{jiang2023mistral} & Open source by Mistral AI & STELLA-ESM3-Mistral-7B-Instruct-v0.2 \\
    & Phi-3-mini-128k-instruct~\citep{abdin2024phi3technicalreporthighly} & Open source by Microsoft & STELLA-ESM3-Phi-3-mini-128k-instruct \\
    & BioMistral-7B-DARE \tnote{a}\; & Tailored for biomedical domain  & STELLA-ESM3-BioMistral-7B-DARE \\
    & BioMedGPT-LM-7B \tnote{b}\; \cite{luo2023biomedgpt} & Tailored for biomedical domain & STELLA-ESM3-BioMedGPT-LM-7B \\
    
    \midrule
    \multirow{5}{*}{Prot2Text}
    & Llama-3.1-8B-Instruct & Open source by Meta & STELLA-Prot2Text-Llama-3.1-8B-Instruct \\
    & Llama-3-8B-Instruct & Open source by Meta & STELLA-Prot2Text-Llama-3-8B-Instruct \\
    & Mistral-7B-Instruct-v0.2 & Open source by Mistral AI & STELLA-Prot2Text-Mistral-7B-Instruct-v0.2 \\
    & Phi-3-mini-128k-instruct & Open source by Microsoft & STELLA-Prot2Text-Phi-3-mini-128k-instruct \\
    & BioMistral-7B-DARE & Tailored for biomedical domain & STELLA-Prot2Text-BioMistral-7B-DARE \\
    & BioMedGPT-LM-7B & Tailored for biomedical domain & STELLA-Prot2Text-BioMedGPT-LM-7B \\
    
    \midrule
    \multirow{5}{*}{SaProt}
    & Llama-3.1-8B-Instruct & Open source by Meta & STELLA-SaProt-Llama-3.1-8B-Instruct \\
    & Llama-3-8B-Instruct & Open source by Meta & STELLA-SaProt-Llama-3-8B-Instruct \\
    & Mistral-7B-Instruct-v0.2 & Open source by Mistral AI & STELLA-SaProt-Mistral-7B-Instruct-v0.2 \\
    & Phi-3-mini-128k-instruct & Open source by Microsoft & STELLA-SaProt-Phi-3-mini-128k-instruct \\
    & BioMistral-7B-DARE & Tailored for biomedical domain & STELLA-SaProt-BioMistral-7B-DARE \\
    & BioMedGPT-LM-7B & Tailored for biomedical domain  & STELLA-SaProt-BioMedGPT-LM-7B \\
    
    \bottomrule
    \end{tabular}

    \begin{tablenotes}
        \item[a] Merge~\citep{yu2024language} of Mistral-7B-Instruct-v0.1 and BioMistral-7B~\citep{labrak2024biomistral} which was further pre-trained on top of Mistral-7B-Instruct-v0.1 using PubMed Central Open Access from https://www.ncbi.nlm.nih.gov/pmc/tools/openftlist/
        \item[b] Increamtally pre-training from Llama-2-7B-Chat with S2ORC~\citep{lo-wang-2020-s2orc} corpus.
    \end{tablenotes}
    
    \end{threeparttable}
    }
    \vspace{-0.5cm}
\end{table*}

\section{Ablation of Training Epochs with Hybrid Datasets}\label{apx:ablation_training_epochs_trend}

\begin{figure*}[!ht]
    \centering
    \includegraphics[width=\linewidth]{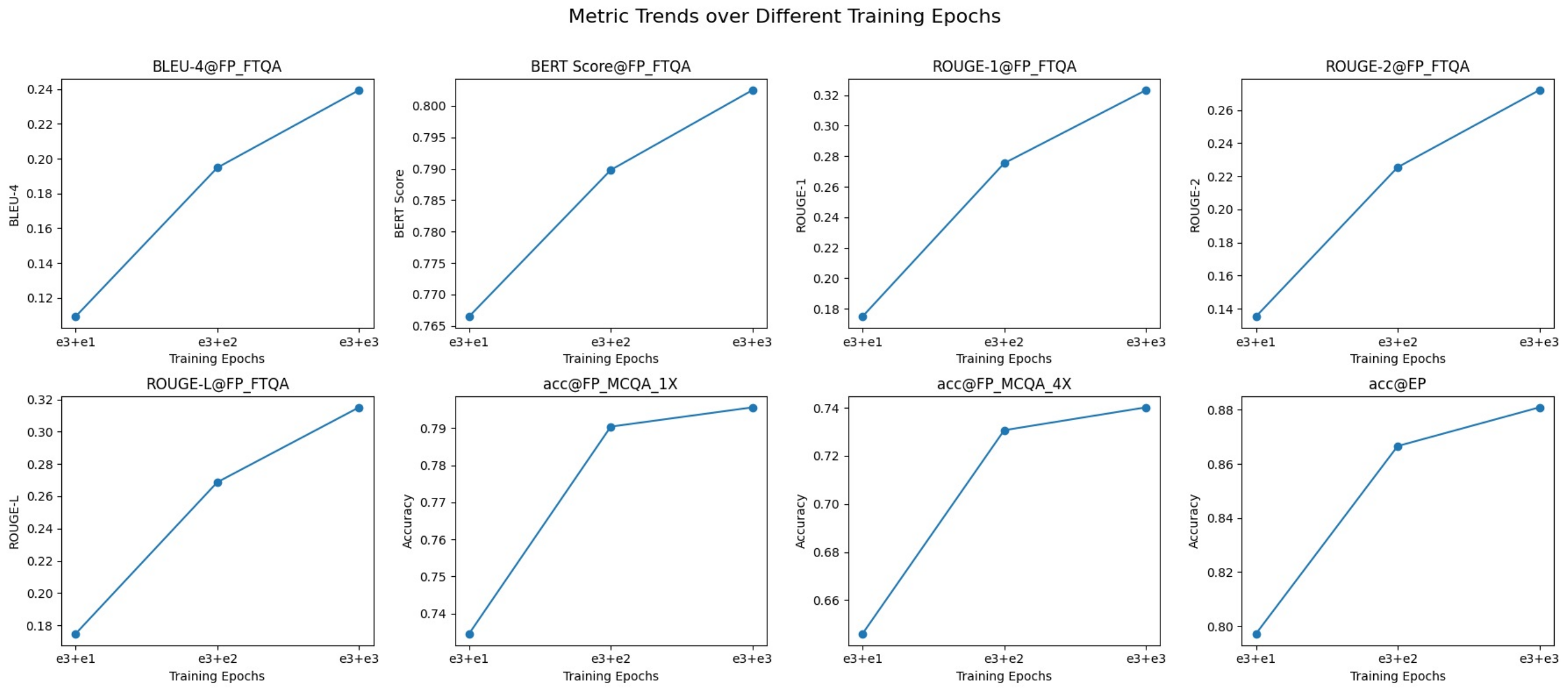}
    \caption{Metrics trend for training with the dataset mix3 over different training epochs.}
    \label{fig:ablation_training_epochs_trend}
\end{figure*}

An ablation study was conducted to evaluate model performance across varying training epochs. For the training with the hybrid three training datasets, i.e., \texttt{Func\_ft\_train}, \texttt{Func\_mc\_train} and \texttt{Enzyme\_train}, all metrics demonstrated consistent improvement with extended training, progressing from (e3+e1) to (e3+e3), as illustrated in Figure~\ref{fig:ablation_training_epochs_trend}. This trend underscores the positive effect of prolonged training on model performance and emphasizes the significance of appropriate training duration to optimize predictive performance. Each subfigure in Figure~\ref{fig:ablation_training_epochs_trend} shows how the scores for BLEU-4, BERT Score, ROUGE-1/ROUGE-2/ROUGE-L Scores, and Accuracy change over the training periods labeled as (e3+e1), (e3+e2), and (e3+e3).

\section{\textbf{STELLA in Action: Case Studies of \texttt{FP} Task}}

\begin{figure*}[!htbp]
    \centering
    \includegraphics[width=0.95\textwidth]{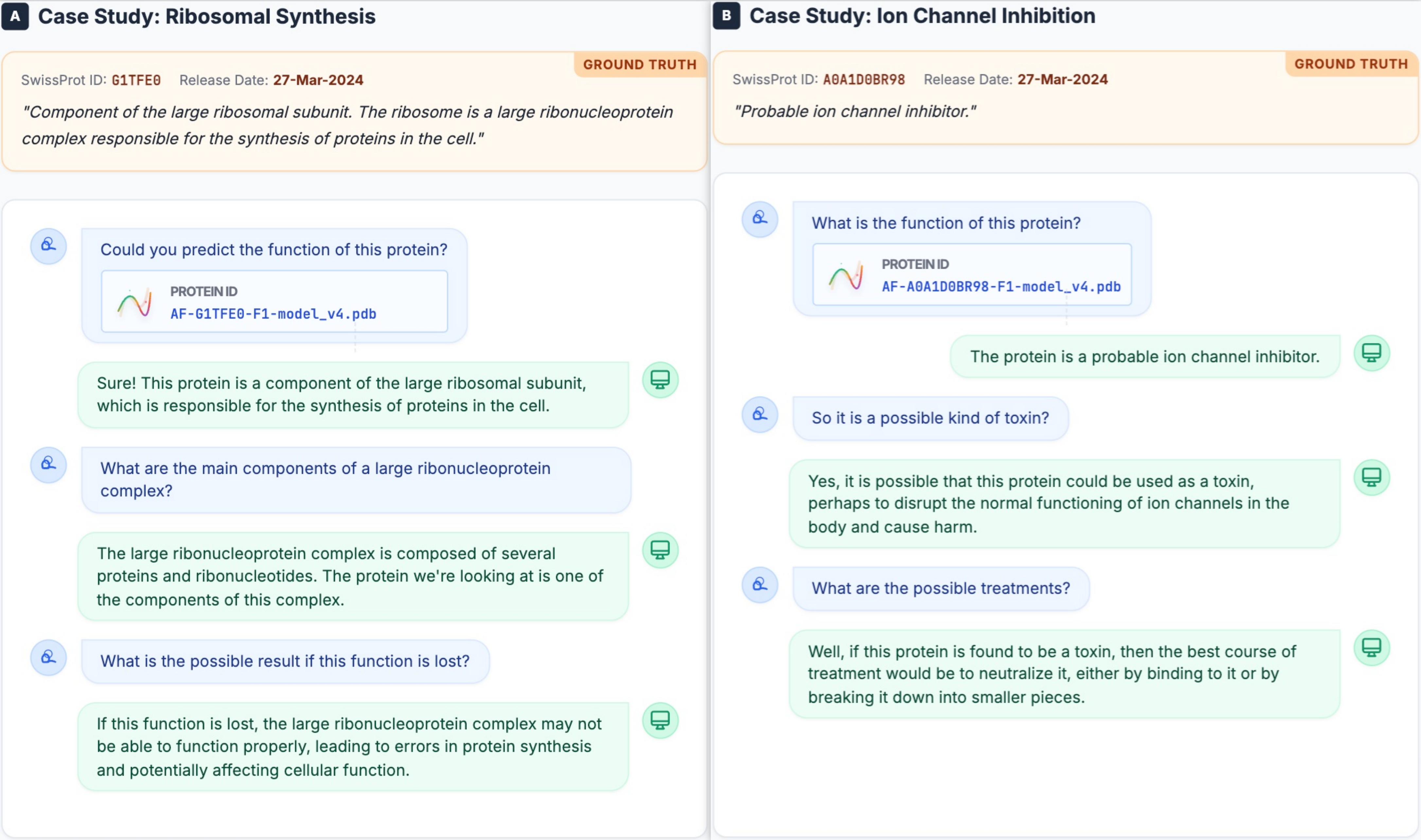}
    \caption{\textbf{Functional annotation examples generated by STELLA-ESM3-Llama-
3.1-8B-Instruct.} The proteins G1TFE0 (left) and A0A1D0BR98 (right) are sourced from Swiss-Prot 2024\_02. The \textcolor{neworange}{\textbf{orange boxes}} indicate ground-truth annotations, while \textcolor{newgreen}{\textbf{green text}} highlights correct and essential functional insights predicted by STELLA.}
    \label{fig:inter_chat1}
    \vspace{-0.05in}
\end{figure*}

STELLA demonstrates feasibility in protein function prediction by integrating sequence-structure represent-ations into LLMs. As illustrated in Figure~\ref{fig:inter_chat1} (left), STELLA excels in following natural language instructions and generating appropriate responses for users. In the example, STELLA correctly identifies the main function—a component of the large ribosomal subunit responsible for the synthesis of proteins in the cell—of a newly reviewed protein G1TFE0 in Swiss-Prot. Additionally, STELLA elaborates on the core constituents of the ribonucleoprotein complex, highlighting its advantage in grasping general knowledge. Furthermore, STELLA showcases its reasoning ability by linking loss of ribosomal function to cellular dysfunctions. 
In Figure~\ref{fig:inter_chat1} (right), STELLA accurately predicts the function of another newly characterized protein in Swiss-Prot, A0A1D0BR98. Upon further inquiry from the user, it explains the details of the toxin mechanisms and provides treatment suggestions. Both examples demonstrate STELLA’s ability in protein function prediction, such as delivering informative, contextually relevant responses to diverse user prompts. Moreover, STELLA shows reasoning ability, which enables it to assist domain experts in uncovering connections between protein functions and disease mechanisms. These results highlight its potential to advance protein biology.


\begin{figure*}[!hbp]
    \vspace{-0.2in}
    \centering
    \includegraphics[width=0.95\textwidth]{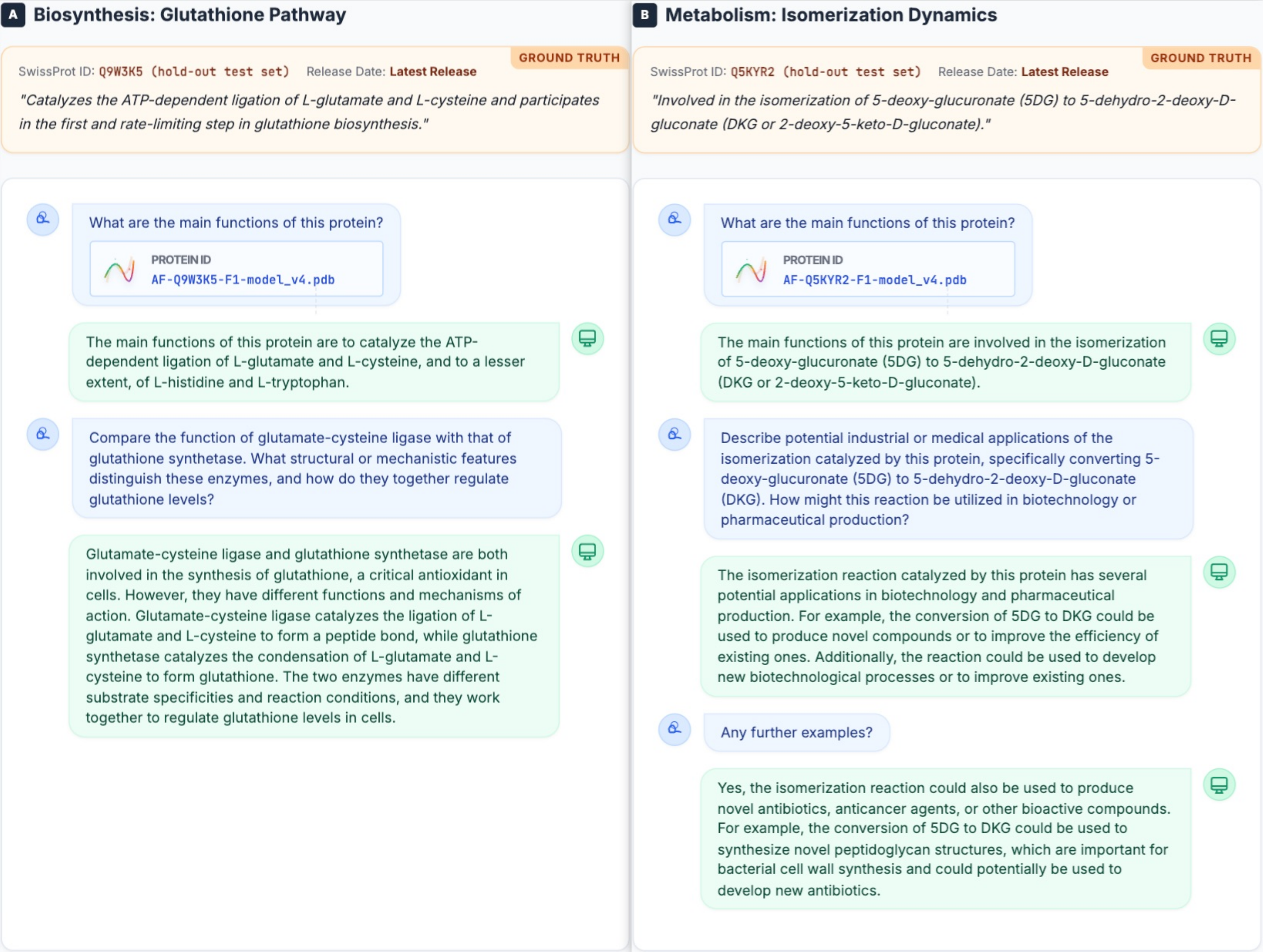}
    \caption{\textbf{Functional annotation examples generated by STELLA-ESM3-Llama-
3.1-8B-Instruct.} The proteins Q9W3K5 (left) and Q5KYR2 (right) are sourced from the hold-out test set. The \textcolor{neworange}{\textbf{orange boxes}} indicate ground-truth annotations, while \textcolor{newgreen}{\textbf{green text}} highlights correct and essential functional insights predicted by STELLA.}
    \label{fig:inter_chat2}
\end{figure*}

In addition, Figure~\ref{fig:inter_chat2} presents two representative case studies demonstrating the capabilities of STELLA-ESM3-Llama-3.1-8B-Instruct in exploring and predicting protein functions as well as other biologically relevant properties. In these examples, STELLA behaves in a step-by-step conversational manner to respond to user prompts, highlighting its ability to reason over protein information to provide accurate and biologically meaningful explanations.

\section{Data Augmentation for \texttt{Function} Dataset}\label{apx:data_aug_methods}
The purpose of enriching the \texttt{Func\_ft\_train} dataset into \texttt{Func\_ft\_train\_aug} is specifically to enhance the conversational capabilities of our model. The motivation behind this data augmentation is to transform ground truth functional descriptions stored in databases into dialogues, thus preserving scientific accuracy as well as natural conversational interactions. The augmentation process involves the following main steps.

\textbf{1.} Prepare ground truth functional descriptions as LLM input: We start with accurate, expert-reviewed descriptions of protein functions. For example: "Required for accurate and efficient protein synthesis under certain stress conditions. May act as a fidelity factor of the translation reaction by catalyzing a one-codon backward translocation of tRNAs on improperly translocated ribosomes. Back-translocation proceeds from a post-translocation (POST) complex to a pre-translocation (PRE) complex, thus giving elongation factor G a second chance to translocate the tRNAs correctly. Binds to ribosomes in a GTP-dependent manner."\footnote{\url{https://www.uniprot.org/uniprotkb/O67618/entry}}

\textbf{2.} Prompt Llama-2-13B-Chat, which maintains computational efficiency while ensuring diversity, to generate conversational data: We utilize the Llama-2-13B-Chat model to convert these structured descriptions into conversational question-answer pairs. Specifically, we employ the following prompt to ensure detailed and meaningful dialogues: "\textit{Given a functional description of the protein, design two or three rounds of questions and answers based on this description. Ensure the content is detailed. The output format is: ['Q':, 'A':, 'Q':, 'A':].}"

\textbf{3.} Save the augmentated data in the format as shown in the example Box \ref{box:ex_func_ft_train_aug} in Appexdix~\ref{apx:examples_instruction_data}.

\section{Computational Cost}
Briefly, STELLA was trained on 8x NVIDIA A100 GPUs for approximately 51 hours total. This includes Stage 1 ($\sim$10.5 hours for 3 epochs) and Stage 2 ($\sim$40.5 hours for 6 epochs). Due to the linear adapter and frozen-encoder strategy, STELLA is significantly more scalable than models requiring full GNN training or complex cross-attention layers.

\section{Diversified Instructions for Training Data Construction}\label{detailed_instructions}
This section presents some exampels of the diversified natural language instructions (see Box~\ref{box:func_instructions_1}-\ref{box:enzyme_instructions_1}) generated by ChatGPT, designed for two tasks--\texttt{FP} and \texttt{EP}. 


\begin{mybox}[label={box:func_instructions_1}]{Diversified instructions of the protein functional prediction task}
\begin{itemize}[]
    \item May I request an elaborate overview of the function linked to the protein?
    \item May I request a detailed overview of the function associated with the protein?
    \item May I request a detailed breakdown of the function associated with the protein?
    \item ...
\end{itemize}
\end{mybox}

\begin{mybox}[label={box:enzyme_instructions_1}]{Diversified instructions of the enzyme-catalyzed reaction prediction task}
\begin{itemize}
    \item Could you provide the enzyme name used for this protein?
    \item Can you share the enzyme name linked to this protein referred to as?
    \item What's the enzyme name titled for this protein?
    \item ...
\end{itemize}
\end{mybox}

\section{Data Examples of OPI-Struc}\label{apx:examples_instruction_data}
This section lists examples (see Box~\ref{box:ex_func_ft_train}-\ref{box:ex_func_mc_train}) from the OPI-Struc dataset, which supports multimodal instruction tuning for STELLA. Each example pairs a protein entry with a diverse natural language instruction and its corresponding output. The examples span various task types, reflecting the dataset’s comprehensive coverage and the model's instruction-following ability in biological contexts.

\begin{mybox}[label={box:ex_func_ft_train_aug},top=1mm, bottom=1mm, left=1mm, right=1mm]{An example of \texttt{Func\_ft\_train\_aug} dataset}

\begin{lstlisting}[breaklines=true,basicstyle=\ttfamily\small]
[
  {
    "swissprot_id": "Q1WT60",
    "sequence": "MENKKLHIALLFGGN...YDFTKLGTEKVGKKIIGE",
    "conversations": [
      {
        "from": "human",
        "value": "What is the main function of this protein?\n<structure>"
      },
      {
        "from": "gpt",
        "value": "The main function of this protein is cell wall formation."
      },
      {
        "from": "human",
        "value": "Can you explain the process of cell wall formation?"
      },
      {
        "from": "gpt",
        "value": "Certainly! The protein helps to synthesize and modify the cell wall, which is a critical structure that provides support, protection, and cell-to-cell interaction. The cell wall is composed of various components, including polysaccharides, proteins, and other molecules. Our protein plays a key role in the synthesis and modification of these components to form a strong and stable cell wall."
      },
      {
        "from": "human",
        "value": "How does the protein contribute to the synthesis of the cell wall?"
      },
      {
        "from": "gpt",
        "value": "The protein helps to catalyze the formation of glycosidic bonds between the cell wall components. These bonds are crucial for the stability and integrity of the cell wall. Additionally, the protein can also modify existing glycosidic bonds to further strengthen the cell wall."
      }
    ],
    "AFDB_id": "AF-Q1WT60-F1-model_v4"
  },
]
\end{lstlisting}
\end{mybox}

\begin{mybox}[label={box:ex_enzyme_train},top=1mm, bottom=1mm, left=1mm, right=1mm]{An example of \texttt{Enzyme\_train} dataset}
\begin{lstlisting}[breaklines=true,basicstyle=\ttfamily\small]
[
  {
    "PDB_id": "5ivp.A",
    "conversations": [
      {
        "from": "human",
        "value": "Could you inform me of the enzyme name tagged to this protein?<structure>\n"
      },
      {
        "from": "gpt",
        "value": "aminoacyl-tRNA hydrolase"
      }
    ]
  },
]
\end{lstlisting}
\end{mybox}

\vspace{-0.25cm}
\begin{mybox}[label={box:ex_func_mc_train}, top=1mm, bottom=1mm, left=1mm, right=1mm]{An example of \texttt{Func\_mc\_train} dataset}

\begin{lstlisting}[breaklines=true,basicstyle=\ttfamily\small]
[
  {
    "swissprot_id": "P62877",
    "sequence": "MAAAMDVDTPSGTNS...RQVCPLDNREWEFQKYGH",
    "conversations": [
      {
        "from": "human",
        "value": "<structure>\n
        What are the main functions of this protein?\n
        A. E3 ubiquitin ligase component of multiple cullin-RING-based E3 ubiquitin-protein ligase (CRLs) complexes which mediate the ubiquitination and subsequent proteasomal degradation of target proteins, including proteins involved in cell cycle progression, signal transduction, transcription and transcription-coupled nucleotide excision repair. CRLs complexes and ARIH1 collaborate in tandem to mediate ubiquitination of target proteins, ARIH1 mediating addition of the first ubiquitin on CRLs targets. The functional specificity of the E3 ubiquitin-protein ligase complexes depends on the variable substrate recognition components. As a component of the CSA complex promotes the ubiquitination of ERCC6 resulting in proteasomal degradation. Recruits the E2 ubiquitin-conjugating enzyme CDC34 to the complex and brings it into close proximity to the substrate. Probably also stimulates CDC34 autoubiquitination. May be required for histone H3 and histone H4 ubiquitination in response to ultraviolet and for subsequent DNA repair. Promotes the neddylation of CUL1, CUL2, CUL4 and CUL4 via its interaction with UBE2M. Involved in the ubiquitination of KEAP1, ENC1 and KLHL41. In concert with ATF2 and CUL3, promotes degradation of KAT5 thereby attenuating its ability to acetylate and activate ATM.\n
        B. Part of the MIS12 complex which is required for normal chromosome alignment and segregation and kinetochore formation during mitosis.\n
        C. Catalyzes the cyanide-resistant oxidation of ubiquinol and the reduction of molecular oxygen to water, but does not translocate protons and consequently is not linked to oxidative phosphorylation. May increase respiration when the cytochrome respiratory pathway is restricted, or in response to low temperatures.\n
        D. Pair-rule protein expressed in a segmentally repeating pattern to define the polarity of embryonic segments. Capable of sequence-specific DNA-binding.\n
        Answer with the option's letter from the given choices directly."
      },
      {
        "from": "gpt",
        "value": "A"
      }
    ],
    "AFDB_id": "AF-P62877-F1-model_v4"
  },
]
\end{lstlisting}
\end{mybox}

\begin{mybox}[label={box:ex_func_ft_train},top=1mm, bottom=1mm, left=1mm, right=1mm]{An example of the \texttt{Func\_ft\_train} dataset}
\begin{lstlisting}[breaklines=true,basicstyle=\ttfamily\small]
{
  "swissprot_id": "Q0BWM9",
  "sequence": "MFNKQSVSLEWAGRT...DQETGKELAEEAGDDASED",
  "conversations": [
    {
      "from": "human",
      "value": "Can you furnish a comprehensive description outlining the function associated with the protein?\n<structure>"
    },
    {
      "from": "gpt",
      "value": "Involved in mRNA degradation. Catalyzes the phosphorolysis of single-stranded polyribonucleotides processively in the 3'- to 5'-direction."
    }
  ],
  "AFDB_id": "AF-Q0BWM9-F1-model_v4"
}
\end{lstlisting}
\end{mybox}

\begin{table*}[p] 
  \centering
  \renewcommand{\arraystretch}{1.1}
  
  \begin{minipage}{\textwidth}
    \section{Hold-out Evaluation Results on the \texttt{FP\_ft\_eval} Benchmark}\label{apx:results_hold_out_eval}
    Table \ref{apx:tab:results_hold_out_eval} presents a comparative analysis of STELLA against baselines and state-of-the-art methods using the \texttt{FP\_ft\_test} hold-out set. The results underscore STELLA's capability in generating high-fidelity functional descriptions.
    
    \vspace{1em}
    \centering
    \captionof{table}{\textbf{Hold-out evaluation of FP performance.} \textbf{Bold}: best; \underline{underline}: runner-up.}
    \label{apx:tab:results_hold_out_eval}
    \small
    \begin{tabularx}{0.98\textwidth}{l YYYYY} 
      \toprule
       \multirow{2}[2]{*}{Model/Method} 
      & \multirow{2}[2]{*}{BLEU-4 $ \uparrow $} 
      & \multirow{2}[2]{*}{BERTScore $ \uparrow $} 
      & \multicolumn{3}{c}{ROUGE Score $ \uparrow $} \\
      \cmidrule(r){4-6}
      & & & ROUGE-1 & ROUGE-2 & ROUGE-L \\ 
      \midrule
       Foldseek & 0.3627 & 0.8358 & 0.4799 & 0.4027 & 0.4586 \\
       Prot2Text$_{BASE}$ & 0.3511 & 0.8430 & 0.5059 & 0.4271 & 0.4849 \\
       Prot2Text$_{LARGE}$ & 0.3629 & \underline{0.8520} & \underline{0.5368} & \underline{0.4560} & \underline{0.5140} \\
       ProteinChat & 0.1918 & 0.7970 & 0.3957 & 0.2799 & 0.3648 \\
       STELLA (e3+e6) & \textbf{0.4300} & \textbf{0.8564} & \textbf{0.5423} & \textbf{0.4747} & \textbf{0.5257} \\
      \bottomrule
    \end{tabularx}
  \end{minipage}

  \vspace{3em} 

  \begin{minipage}{\textwidth}
    \section{Zero-shot Temporal Evaluation on the \texttt{FP\_ft\_eval\_v2401} Benchmark}\label{apx:results_zero_shot_eval}
    Table \ref{apx:tab:results_zero_shot_eval} illustrates the zero-shot Out-of-Distribution (OOD) generalization of STELLA on the \texttt{FP\_ft\_eval\_v2401} benchmark. This temporal test set assesses the model's generalization on proteins characterized after the training data cutoff.

    \vspace{1em}
    \centering
    \captionof{table}{\textbf{Zero-shot temporal OOD evaluation.} \textbf{Bold}: best; \underline{underline}: runner-up.}
    \label{apx:tab:results_zero_shot_eval}
    \small
    \begin{tabularx}{0.98\textwidth}{l YYYYY} 
      \toprule
      \multirow{2}[2]{*}{Model} 
      & \multirow{2}[2]{*}{BLEU-4 $ \uparrow $} 
      & \multirow{2}[2]{*}{BERTScore $ \uparrow $} 
      & \multicolumn{3}{c}{ROUGE Score $ \uparrow $} \\  
      \cmidrule(r){4-6}
      & & & ROUGE-1 & ROUGE-2 & ROUGE-L \\ 
      \midrule
      STELLA-ESM3-Llama-3.1-8B-Instruct        & \underline{0.0489} & 0.7565 & 0.2210 & \textbf{0.1085} & 0.1867 \\
      STELLA-Prot2Text-Llama-3.1-8B-Instruct   & 0.0425 & 0.7555 & 0.2454 & 0.1020 & \underline{0.1919} \\
      STELLA-Prot2Text-Llama-3-8B-Instruct     & \textbf{0.0510} & \underline{0.7605} & \underline{0.2486} & \underline{0.1062} & 0.1918 \\
      STELLA-Prot2Text-Mistral-7B-Instruct-v0.2 & 0.0440 & \textbf{0.7685} & \textbf{0.2529} & 0.1046 & \textbf{0.1975} \\
      ProteinChat & 0.0205 & 0.7413 & 0.2121 & 0.0855 & 0.1691 \\
      \bottomrule
    \end{tabularx}
  \end{minipage}

  \vspace{3em}

  \begin{minipage}{\textwidth}
    \section{Comparative Analysis of the Stage-wise Training Strategy}\label{apx:ablation_training_stages}
    Table \ref{apx:tab:ablation_training_stages} compares the efficacy of the decoupled training strategy. By comparing single-stage and two-stage training for STELLA-ESM3-Llama-3.1-8B-Instruct, we highlight the benefit of two-stage optimization in multimodal instruction tuning.

    \vspace{1em}
    \centering
    \captionof{table}{\textbf{Ablation of the stage-wise training strategy on the FP\_ft\_eval benchmark.} S1: Stage 1, S2: Stage 2. \textbf{Bold}: best.}
    \label{apx:tab:ablation_training_stages}
    \small
    \begin{tabularx}{0.98\textwidth}{ccc YYYYY}
      \toprule
      \multirow{2}[2]{*}{Strategy} 
      & \multirow{2}[2]{*}{S1 Epoch} 
      & \multirow{2}[2]{*}{S2 Epoch} 
      & \multirow{2}[2]{*}{BLEU-4 $ \uparrow $} 
      & \multirow{2}[2]{*}{BERTScore $ \uparrow $} 
      & \multicolumn{3}{c}{ROUGE Score $ \uparrow $} \\
      \cmidrule(r){6-8}
      & & & & & ROUGE-1 & ROUGE-2 & ROUGE-L \\ 
      \midrule    
       Single-stage & - & e1 & 0.2233 & 0.7885 & 0.3530 & 0.2631 & 0.3350 \\
       Single-stage & - & e2 & 0.3099 & 0.8199 & 0.4346 & 0.3522 & 0.4160 \\
       Single-stage & - & e3 & 0.3642 & 0.8363 & 0.4840 & 0.4073 & 0.4660 \\
      \midrule
       Two-stage & e3 & e1 & 0.2653 & 0.8065 & 0.3938 & 0.3097 & 0.3770 \\
       Two-stage & e3 & e2 & 0.3574 & 0.8363 & 0.4790 & 0.4028 & 0.4617 \\
       Two-stage & e3 & e3 & \textbf{0.4024} & \textbf{0.8496} & \textbf{0.5218} & \textbf{0.4487} & \textbf{0.5041} \\
      \bottomrule
    \end{tabularx}
  \end{minipage}
\end{table*}

\end{document}